\documentclass[aps,pra,reprint,preprintnumbers,showpacs,superscriptaddress]{revtex4-1}

\usepackage{graphicx}
\usepackage{dcolumn}
\usepackage{bm}

\begin{document}

\preprint{Ver.7.1}

\title{Fermi surface reconstruction in FeSe under high pressure}


\author{Taichi Terashima}
\author{Naoki Kikugawa}
\affiliation{National Institute for Materials Science, Tsukuba, Ibaraki 305-0003, Japan}
\author{Andhika Kiswandhi}
\altaffiliation[Present address: ]{Department of Physics, University of Texas at Dallas, Richardson, TX 75080, USA}
\author{David Graf}
\author{Eun-Sang Choi}
\author{James S. Brooks}
\affiliation{National High Magnetic Field Laboratory, Florida State University, Tallahassee, FL 32310, USA}
\author{Shigeru Kasahara}
\author{Tatsuya Watashige}
\author{Yuji Matsuda}
\affiliation{Department of Physics, Kyoto University, Kyoto 606-8502, Japan}
\author{Takasada Shibauchi}
\affiliation{Department of Advanced Materials Science, University of Tokyo, Chiba 277-8561, Japan}
\author{Thomas Wolf}
\author{Anna E. B\"ohmer}
\altaffiliation[Present address: ]{The Ames Laboratory, U.S. Department of Energy, Iowa State University, Ames, Iowa 50011, USA}
\author{Fr\'ed\'eric Hardy}
\author{Christoph Meingast}
\author{Hilbert v. L\"ohneysen}
\affiliation{Institute of Solid State Physics (IFP), Karlsruhe Institute of Technology, D-76021 Karlsruhe, Germany}
\author{Shinya Uji}
\affiliation{National Institute for Materials Science, Tsukuba, Ibaraki 305-0003, Japan}


\date{\today}

\begin{abstract}
We report Shubnikov-de Haas (SdH) oscillation measurements on FeSe under high pressure up to $P$ = 16.1 kbar.
We find a sudden change in SdH oscillations at the onset of the pressure-induced antiferromagnetism at $P$ $\sim$ 8 kbar.
We argue that this change can be attributed to a reconstruction of the Fermi surface by the antiferromagnetic order.
The negative d$T_c$/d$P$ observed in a range between $P$ $\sim$ 8 and 12 kbar may be explained by the reduction in the density of states due to the reconstruction.
The ratio of the transition temperature to the effective Fermi energy remains high under high pressure: $k_BT_c/E_F$ $\sim$ 0.1 even at $P$ = 16.1 kbar.
\end{abstract}


\maketitle


\newcommand{\ud}{\mathrm{d}}
\def\degree{\kern-.2em\r{}\kern-.3em}

\section{Introduction}

Since the discovery of superconductivity (SC) at $T_c$ = 26 K in LaFeAs(O$_{1-x}$F$_x$) by Kamihara \textit{et al.} \cite{Kamihara08JACS}, the iron-based high-$T_c$ materials have been an object of intense research activity.
A recent focus has been on the electronic nematicity and its role in the high-$T_c$ SC \cite{Fernandes14NatPhys}.
In typical iron-pnictide parent compounds such as LaFeAsO or BaFe$_2$As$_2$ \cite{Rotter08PRL, Sasmal08PRL}, a tetragonal-to-orthorhombic structural transition precedes or coincides with stripe-type antiferromagnetic (AFM) order with a wave vector $q$ = ($\pi$, 0) in the so-called one-Fe Brillouin zone.
The origin of the structural transition, often referred to as a nematic transition, is believed to be electronic \cite{Fernandes14NatPhys}.
Both orders are suppressed by pressure or chemical substitution, when SC emerges.
Since the two orders reside in close proximity, it has been argued that spin degrees of freedom rather than orbital ones are the primary cause of the nematic order \cite{Fernandes14NatPhys}.

The compound studied here, FeSe, is a fascinating iron-based superconductor.
Its $T_c$ at ambient pressure is low ($T_c$ $\sim$ 8 K) \cite{Hsu08PNAS}, but the onset of SC can be enhanced up to $\sim$ 37 K by the application of pressure \cite{Mizuguchi08APL, Medvedev09Nmat}.
Moreover, still higher $T_c$ has been reported for single-layer films \cite{Wang12CPL}.
It stands out against the above general view.
It exhibits a structural transition at $T_s$ $\sim$ 90 K but does not order magnetically at ambient pressure \cite{McQueen09PRL}.
Angle-resolved photoemission spectroscopy (ARPES) measurements show a splitting of the $d_{xz}$ and $d_{yz}$ bands below $\sim T_s$ \cite{Tan13NatMat, Shimojima14PRB, Nakayama14PRL}.
NMR measurements find strong spin fluctuations only below $T_s$, providing evidence against spin-driven nematicity \cite{Imai09PRL, Baek14nmat, Bohmer15PRL}.
Note, however, inelastic neutron scattering measurements attest the existence of ($\pi$, 0) AFM fluctuations \cite{Rahn15PRB, Wang15NatMater}. 
$\mu$SR measurements show that AFM order occurs under high pressure \cite{Bendele10PRL, Bendele12PRB}, and recently we have found an anomaly in the temperature dependence of electrical resistance which presumably corresponds to this order \cite{Terashima15JPSJ}.  
The observed pressure-temperature phase diagram (Fig.~\ref{phase}) is at variance with the spin-nematic scenario predicting that the AFM phase is enclosed by the nematic one \cite{Fernandes14NatPhys, Yamase15NJP}. 
On the theoretical side, much effort has been devoted to understand the absence of magnetic order at ambient pressure and the nature of the ground state \cite{Essenberger12PRB, Lischner15PRB, Chubukov15PRB, Mukherjee15PRL, Glasbrenner15NatPhys, Hirayama15JPSJ, Leonov15PRL, Wang15NatPhys, Rong15PRL, Jiang15condmat, Yamakawa15condmat}.

We have previously reported Shubnikov-de Haas (SdH) oscillation measurements on FeSe at ambient pressure \cite{Terashima14PRB}.
Our results have subsequently been confirmed by other groups \cite{Audouard15EPL, WatsonPRB15}.
The observed Fermi surface (FS) deviates significantly from that predicted by band-structure calculations, most likely composed of one hole and one electron cylinder (, although some studies suggest the existence of an additional tiny pocket \cite{Huynh14PRB, Watson15PRL, Zhang15condmat}).
The carrier density is of the order of 0.01 carriers/Fe, one order-of-magnitude smaller than predicted.
Because of the small FS, the effective Fermi energies $E_F$ estimated for the observed orbits are small, and hence the ratios $k_BT_c/E_F$ are large: $k_BT_c/E_F$ = 0.04--0.22.
In single-band superconductors, $k_BT_c/E_F$ = 0.2 would be an indication of the Bardeen-Cooper-Schrieffer (BCS)--Bose-Einstein-condensation (BEC) crossover \cite{Randeria14ARCMP}.
Although the effects of multiband electronic structure on the crossover physics should be scrutinized, the large $k_BT_c/E_F$ ratio may have some relevance to peculiarities of the SC in FeSe \cite{Kasahara14PNAS}.

In this study, we extend SdH measurements to high pressures.
We find a drastic change in SdH oscillations at the onset of the AFM order and argue that this change is due to a reconstruction of the FS by the AFM order.
We suggest that the anomalous reduction in $T_c$ observed in the same pressure range can be attributed to the reduction in the density of states (DOS) due to the reconstruction.
Interestingly, a large ratio of $k_BT_c/E_F$ $\sim$ 0.1 is found at the highest pressure of 16.1 kbar.

\section{Experiments}

We performed four-contact electrical resistance $R$ measurements on FeSe in magnetic fields up to $B$ = 45 T (35 T) at pressures up to $P$ = 16.1 kbar and temperatures down to $T$ = 0.4 K (0.05 K) in Tallahassee.
High-quality single crystals were prepared by a chemical vapor transport method \cite{Bohmer13PRB}: samples A and B were prepared in Karlsruhe, while K in Kyoto.
The electrical contacts were spot-welded, and the low-frequency ac current ($f$ $\sim$ 10--20 Hz) was applied in the $ab$ plane.
The magnetic field was applied in the $c$ direction.
Piston-cylinder type pressure cells made of NiCrAl and BeCu alloys (C\&T Factory, Tokyo) were used \cite{Uwatoko02JPCM}.
The pressure transmitting medium was Daphne 7474 (Idemitsu Kosan, Tokyo), which remains liquid up to 37 kbar at room temperature and assures highly hydrostatic pressure generation in the investigated pressure range \cite{Murata08RSI}. 
The pressure was determined from the resistance variation of calibrated manganin wires.
The applied pressures and the corresponding phase transition temperatures can be seen in Fig.~\ref{phase}.
We measured $R(T)$ at $P$ = 8.9 kbar in sample-K only below $T$ = 16 K, and hence $T_N$ at this pressure is unknown.
The data of Fig.~\ref{phase} suggest that the onset of the AFM coincides with the onset of negative d$T_c$/d$P$ at $P$ $\sim$ 8 kbar.
A recent report by Kaluarachchi \textit{et al.} supports this view: at $P$ = 8.7 kbar, which is very close to the pressure of the local maximum of $T_c$, they have observed the anomaly at $T_N$ slightly above $T_c$ \cite{Kaluarachchi16PRB}.

\section{Results}

\begin{figure}
\includegraphics[width=8.6cm]{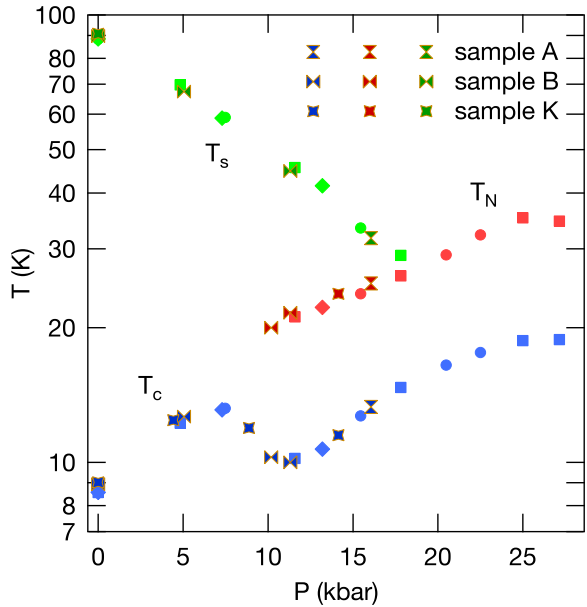}
\caption{\label{phase}(Color online).  High-pressure phase diagram of FeSe. Structural transition at $T_s$, AFM transition at $T_N$ ($T_u$ in \cite{Terashima15JPSJ}), and SC transition at $T_c$.  $T_c$ is defined as the temperature where the resistance becomes zero.  Different symbol shapes correspond to different samples.  Data for samples other than A, B and K are from \cite{Terashima15JPSJ}.}   
\end{figure}

\begin{figure}
\includegraphics[width=8.6cm]{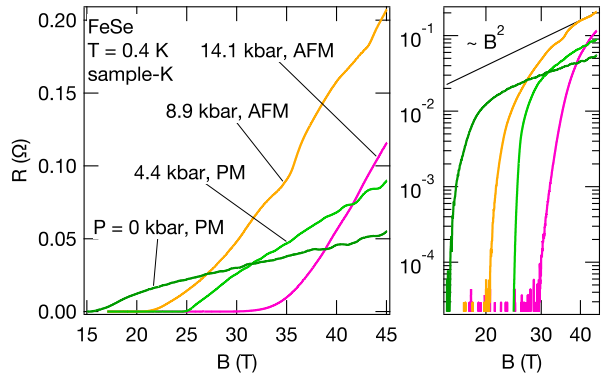}
\caption{\label{RvsB}(Color online).  (Left) Magnetic-field dependence of the resistance in sample K for different pressures. (Right) The same data in a log-log plot.}   
\end{figure}

\begin{figure*}[!t]
\includegraphics[width=8.6cm]{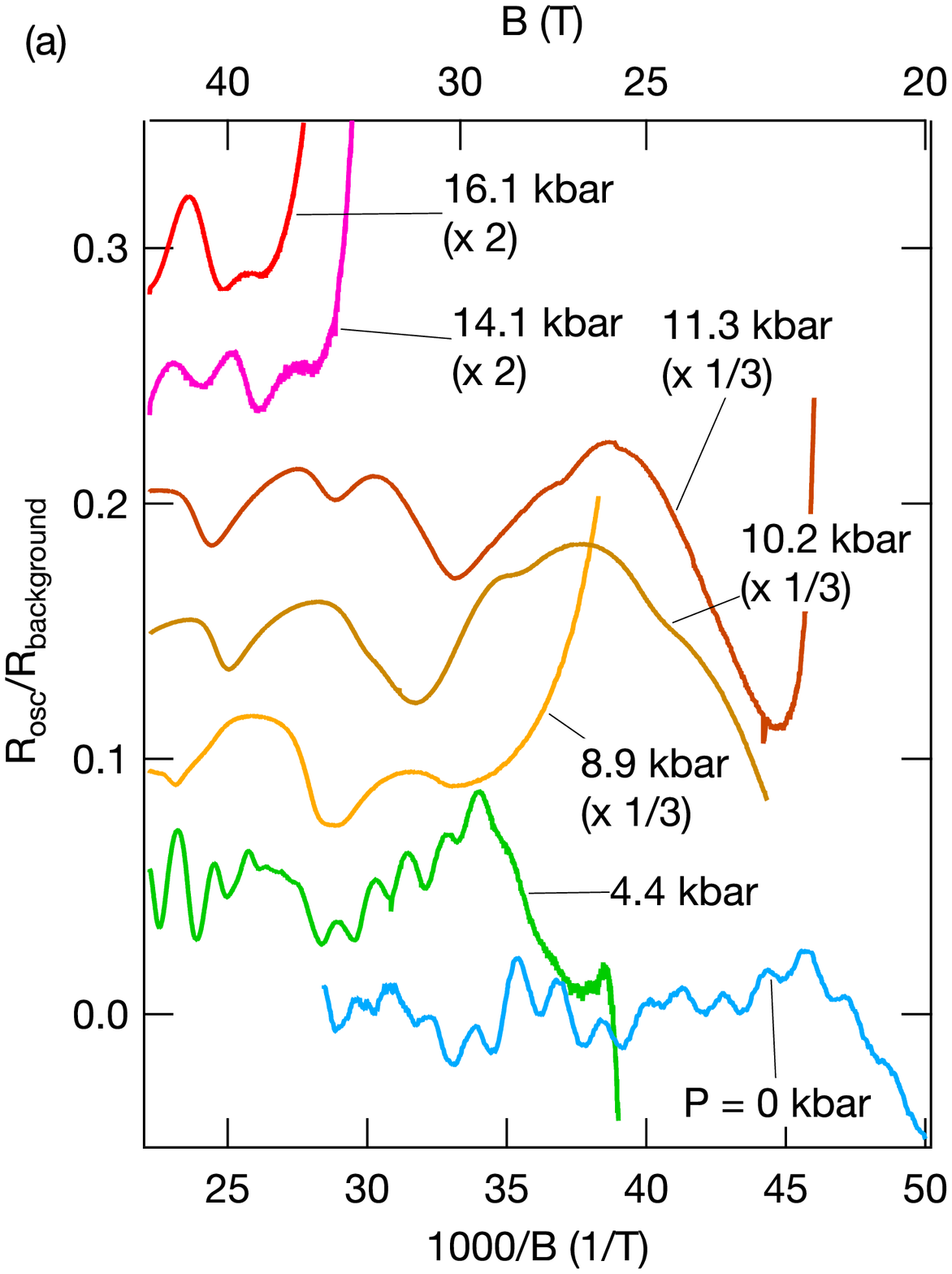}
\includegraphics[width=8.6cm]{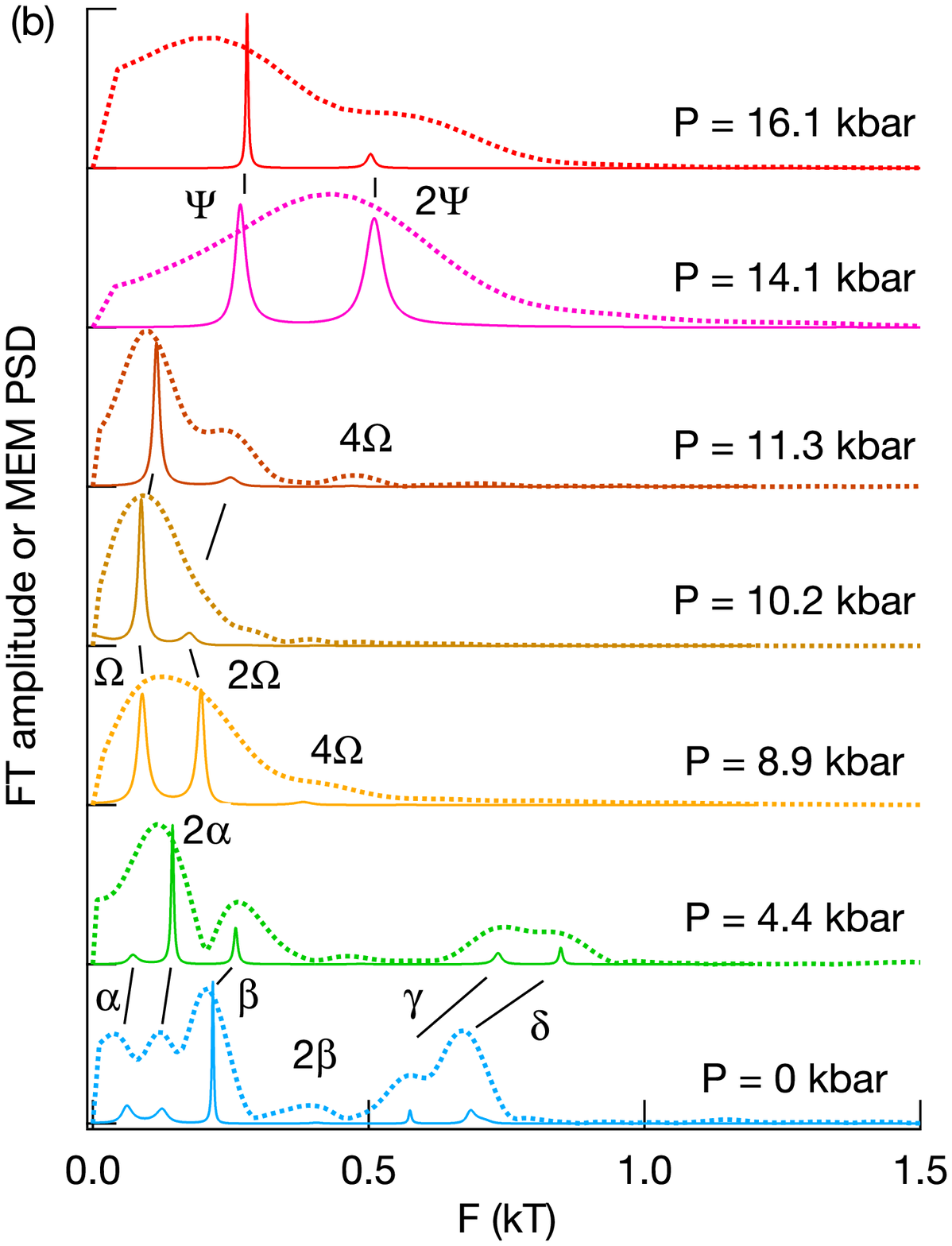}
\caption{\label{Sig_FT}(Color online).  (a) SdH oscillations in FeSe at different pressures as a function of inverse field 1/$B$, and (b) FT amplitudes of the oscillations (dotted lines) and MEM PSDs (solid lines).   [$P$ (kbar), $T$ (K), sample] = (0, 0.046, B), (4.4, 0.38, K), (8.9, 0.36, K), (10.2, 0.43, B), (11.3, 0.64, B), (14.1, 1.1, K), and (16.1, 1.65, A) from bottom to top.  High-$T$ data are shown for high pressures because of lower $B_{c2}$, i.e., wider normal-state field ranges.}   
\end{figure*}

\begin{figure}[!t]
\includegraphics[width=8.6cm]{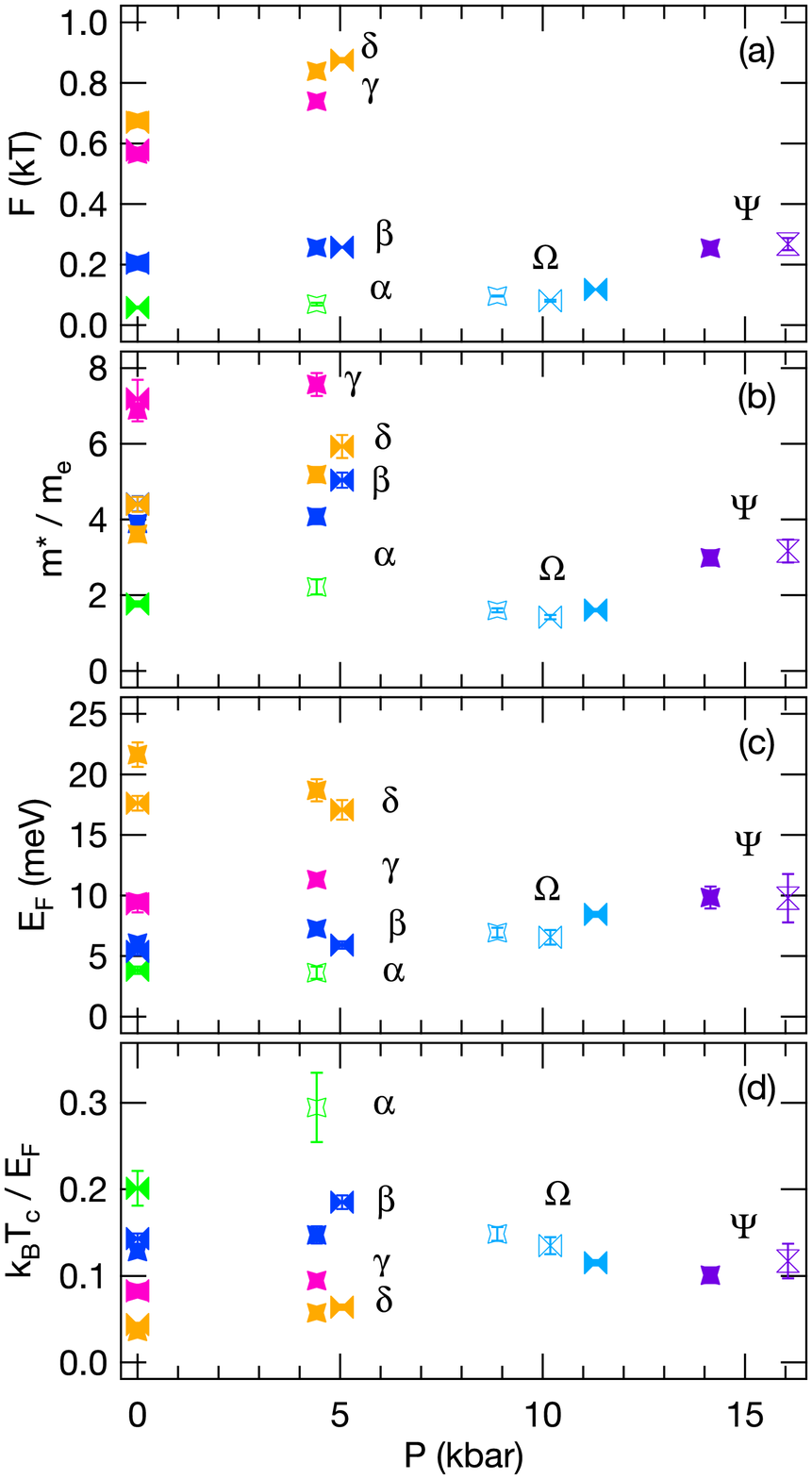}
\caption{\label{param}(Color online).  (a) SdH frequency, (b) effective mass, (c) effective Fermi energy, and (d) $k_BT_c/E_F$ in FeSe as a function of pressure.  Solid and hollow symbols are based on FT and MEM analyses, respectively.  They are consistent with each other and reveal the meaningful pressure evolution of those quantities.  Error bars for $F$ and $m^*$ are based on standard deviations of measurements and fitting errors, respectively.  Errors for the last two quantities are derived from those.  Different symbol shapes correspond to different samples as in Fig.~\ref{phase}.}   
\end{figure}

Figure~\ref{RvsB} shows magnetic-field dependences of the resistance in sample K at $T$ = 0.4 K measured at different pressures.
Within a simple two-carrier model for compensated metals, the transverse magnetoresistance can be expressed as $R(B)=[1+(Ne)^{-2}\sigma_h\sigma_eB^2]R(0)$, where $N$ is the carrier density, $\sigma_{h(e)}$ the hole (electron) conductivity.
The experimental $R(B)$ curves approximately follow the expected $B^2$ dependence at high fields as shown in the log-log plot.
The resistance in the AFM phase at $P$ = 8.9 and 14.1 kbar increases more rapidly with field than that in the paramagnetic (PM) phase at $P$ = 0 and 4.4 kbar, suggesting that in the AFM phase the carrier density $N$ is reduced and/or the conductivity ($\sigma_h$ and/or $\sigma_e$) is enhanced.

We extract SdH oscillations from the measured $R(B)$ curves by subtracting a smoothly varying background, typically modeled by a second- or third-order polynomial.
Figure~\ref{Sig_FT}(a) shows the normalized oscillations $R_{osc}/R_{background}$.
In the PM phase at $P$ = 0 and 4.4 kbar, fast oscillations with periods of the order of $\Delta(1000/B) \sim$1 T$^{-1}$ are clearly visible.
These fast oscillations disappear in the AFM phase at higher pressures.
This sudden change in the SdH oscillations suggests that the FS is reconstructed by the AFM order.

Figure~\ref{Sig_FT}(b) shows the spectral analysis of the SdH oscillations.
We plot Fourier transform (FT) amplitudes (dotted lines) and also power spectral densities (PSD) estimated using the maximum entropy method (MEM) (solid lines) \cite{Press02}.
Because of the increasing upper critical field $B_{c2}$ under pressure (Fig.~\ref{RvsB}), the field range where SdH oscillations can be observed shrinks under pressure [Fig.~\ref{Sig_FT}(a)], which severely degrades the frequency resolution of FT spectra as can be seen in Fig.~\ref{Sig_FT}(b).
In the MEM, an autoregressive model is fitted to data and the spectrum is estimated from its coefficients.
It is known to have high spectral resolution for short data \cite{Ulrych75RGSP, Sigfusson92PRB} and indeed gives sharp spectral peaks for the present data as seen in Fig.~\ref{Sig_FT}(b).
The principle of the MEM and examples of its application to short data are described in Appendix.

The spectrum for $P$ = 0 kbar is consistent with our previous report \cite{Terashima15JPSJ}, where we identified four fundamental frequencies $\alpha$, $\beta$, $\gamma$, and $\delta$ and their harmonics.
The SdH frequency $F$ is proportional to the FS cross-sectional area $A$ normal to the field: $F=\hbar A/(2\pi e)$.
The cross section corresponding to the frequency $\alpha$ is only 0.2\% of the Brillouin zone.
As the pressure is increased in the PM phase to 4.4 kbar, the frequencies increase [see also Fig.~\ref{param}(a)], i.e., the FS expands.

As the sample enters the AFM phase at $P$ = 8.9 kbar, the FT spectrum (dotted curve) shows only a broad low-frequency peak with a shoulder.
The MEM PSD (solid curve) indicates that the broad peak is actually superposition of a fundamental, labeled $\Omega$, and its second harmonic 2$\Omega$.
A weak peak appears near $F \sim$0.4 kT in the MEM PSD.
It can be attributed to the fourth harmonic 4$\Omega$ and can explain the weak shoulder of the FT spectrum.
Although the fundamental frequency $\Omega$ has less than two oscillation periods in the experimental field range, the observation of the harmonics corroborates its existence.
There is no high-frequency fundamental comparable in size to $\gamma$ or $\delta$, confirming the conclusion (i.e, the disappearance of the high frequencies) drawn above from the oscillation signal itself.
At $P$ = 10.2 and 11.3 kbar, the existence of the fundamental frequency $\Omega$ near $F \sim$0.1 kT becomes more unambiguous.
A peak, albeit broad, develops near the position of $\Omega$ even in the FT spectra [Fig.~\ref{Sig_FT}(b)].
Further, the fundamental oscillation can directly be identified in the oscillation signals themselves [Fig.~\ref{Sig_FT}(a)]:
At both pressures, the peaks are observed roughly at $1000/B \sim28$ and 38 T$^{-1}$, and the interval between them corresponds to one oscillation period.
Fine structures, especially the doublet structure near $1000/B \sim28$ T$^{-1}$ at $P$ = 11.3 kbar, might be attributed to the spin splitting, i.e., splitting of the contributions of up- and down-spin electrons \cite{Shoenberg84}.
The above three pressures are in the pressure region where d$T_c$/d$P$ $<$ 0 (Fig.~\ref{phase}).
At $P$ = 14.1 and 16.1 kbar, where d$T_c$/d$P$ $>$ 0, a fundamental frequency appears near $F \sim$0.3 kT with the second harmonic [Fig.~\ref{Sig_FT}(b)].
Since the frequency is more than two times larger than $\Omega$, it is unclear whether this frequency has the same origin as $\Omega$ and hence we label it $\Psi$.
There is still no high-frequency fundamental.

Effective masses associated with the cyclotron orbits can be estimated from the temperature dependences of the oscillation amplitudes \cite{Shoenberg84}.
Since the frequencies $\alpha$, $\Omega$, and $\Psi$ are low, we can observe only less than two oscillation periods in the field ranges available for the mass analysis.
We therefore use the amplitudes of the second harmonics to estimate the associated masses for those frequencies.
We use FT amplitudes as far as possible (solid symbols in Fig.~\ref{param}), but when the frequency resolution is not sufficient we use MEM PSD (hollow symbols in Fig.~\ref{param}).
In the latter case, the square root of the integrated power under a frequency peak is assumed to be an oscillation amplitude for the frequency \cite{Ulrych75RGSP}, as described in Appendix.
The estimated masses are shown in Fig.~\ref{param}(b).
Note that the frequencies and effective masses estimated from the FT and MEM analyses, shown by solid and hollow symbols, respectively, are consistent with each other and reveal the meaningful pressure evolution of those quantities.
The effective Fermi energy $E_F$ can be estimated from $F$ and $m^*$ for each of the observed orbits, using the following formulae: $E_F= \hbar^2 k_F^2/(2m^*)$, and $A=\pi k_F^2$ [Fig.~\ref{param}(c)].
The ratio of $T_c$ and $E_F$ is shown in Fig.~\ref{param}(d).
We note that the ratio remains high at high pressures.

\section{Discussion}

We first discuss the pressure dependence of the frequencies and effective masses in the PM phase.
Naively, the pressure would increase the band width, which would result in increase in FS sizes via increased band overlap and decrease of the effective masses.
The former is consistent with the increase in $F$ from P = 0 to 4.4 kbar in the PM phase [Fig.~\ref{param}(a)].
The $\gamma$ and $\delta$ frequencies increase more rapidly than $\alpha$ and $\beta$.
This indicates an increased three-dimensionality with pressure, as the $\gamma$ and $\delta$ ($\alpha$ and $\beta$) frequencies are attributed to maximum (minimum) cross sections normal to the $c$ axis of FS cylinders \cite{Terashima15JPSJ}.
On the other hand, the effective masses do not decrease with pressure, but rather exhibit a moderately increasing trend [Fig.~\ref{param}(b)].
This may indicate that the mass enhancement due to spin fluctuations gets stronger with pressure as suggested by NMR measurements under high pressure \cite{Imai09PRL}.
In BaFe$_2$(As$_{1-x}$P$_x$)$_2$, the existence of an AFM quantum critical point has been argued on the basis of the diverging effective mass towards the onset of the AFM order \cite{Shishido10PRL, Hashimoto12Sci, Walmsley13PRL, Shibauchi14ARCMP}.
In the present case, the data points are not sufficient to deduce a divergence of the effective masses, and further investigations are necessary. 

We now discuss the sudden change in the SdH oscillations at the onset of the AFM order.
Our view is that it is due to the reconstruction of the Fermi surface by the AFM order, as already stated above.
In the following we argue that this view is the most rational by considering possible objections to it.

First of all, we can rule out the possibility that the disappearance of the high frequencies at $P$ = 8.9 kbar (Fig.~\ref{Sig_FT}) is due to the pressure inhomogeneity or non-hydrostaticity.
The pressure medium used in the present study, Daphne 7474, remains liquid up to 37 kbar at room temperature \cite{Murata08RSI}.
A study using the ruby fluorescence technique in a diamond anvil cell has experimentally confirmed the hydrostatic pressure generation up to this solidification pressure at room temperature \cite{Klotz09JPhysD}.   
Further, another study using a similar method has confirmed that a high level of the hydrostaticity is maintained at 77 K \cite{Tateiwa09RSI}:  
The width of the ruby $R_1$ line, a measure of the pressure inhomogeneity, at 20 kbar is approximately two times smaller than that in the case of Flourinert, a frequently used pressure medium, and nearly the same as that in He.
In addition, the line width shows only a very weak pressure dependence up to 20 kbar, and there is no rapid change in a pressure region between 4.4 and 8.9 kbar that could explain the sudden disappearance of the high frequencies at $P$ = 8.9 kbar.

The high level of the hydrostaticity has also been demonstrated in previous high-pressure quantum oscillations measurements using Daphne 7474 as a pressure transmitting medium.
Note that higher-frequency oscillations are generally more susceptible to the pressure inhomogeneity, because the corresponding cyclotron orbits in real space are larger.
In the case of CeRhSi$_3$, for example, high frequencies up to $F \sim$22 kT, more than one order-of-magnitude larger than the present $\gamma$ and $\delta$ frequencies, have been observed at about 30 kbar \cite{Terashima07PRB}.
In KFe$_2$As$_2$, which is approximately as soft and easy to cleave as FeSe and hence can be a good reference compound for comparison, frequencies in a range $F$ = 2 -- 3 kT have continuously been observed from $P$ = 0 to 24.7 kbar with similar amplitudes at all pressures \cite{Terashima14PRB_KFA}.
These clearly indicate that it is difficult to attribute the present disappearance of the high frequencies at $P$ = 8.9 kbar to the pressure inhomogeneity.

One might argue that, instead of a FS reconstruction, phase separation might occur:
AFM domains are phase-separated from PM/SC domains, and the $\Omega$ and $\Psi$ frequencies originate from the latter and are the same as the $\alpha$ frequency.
In this view, the other frequencies disappear simply because of increased scattering.
This scenario might sound especially plausible since the $\mu$SR measurements show that the AFM order is partially suppressed by the onset of SC in the pressure region where d$T_c$/d$P$ $<$ 0 \cite{Bendele10PRL, Bendele12PRB}.
Note however that SC is, in fact, killed by the magnetic field needed to observe SdH oscillations.
We further argue against this phase-separation scenario as follows:
The pressure evolution of the frequencies and effective masses of $\alpha$, $\Omega$, and $\Psi$ can not be explained by simple extrapolation of the behavior of $F$ and $m^*$ in the PM phase, which would be monotonic increase in both quantities [Fig.~\ref{param}(a) and (b)].
Secondly, the oscillation amplitude of $\Omega$ is much larger than that of $\alpha$: note that the oscillations at $P$ = 8.9, 10.2, and 11.3 kbar in Fig.~\ref{Sig_FT}(a) are scaled down by a factor of 1/3.
This suggests that scattering is reduced at those pressures, and hence it is difficult to explain the disappearance of the other frequencies by increased scattering.
Lastly, if there is no reconstruction, the enhanced magnetoresistance in the AFM phase (Fig.~\ref{RvsB}) requires an enhancement of $\sigma_h$ or $\sigma_e$ since $N$ remains nearly the same [see the above expression for $R(B)$].
This once more contradicts the assumption of increased scattering.

One might still wonder if the tiny FS of FeSe could be shared by the SC and AFM.
It has been argued that the AFM order in the iron-based high-$T_c$ compounds is quite different from the spin-density-wave order in chromium, a weakly correlated metal, and that the electrons have both itinerant and localized character \cite{Dai12NatPhys}.
More specifically, the stabilization of the AFM order is not provided by FS nesting alone but involves electrons more than 1 eV below the Fermi level \cite{Johannes09PRB, Heil14PRB}.
It therefore seems possible that the AFM order in FeSe is stabilized without gapping large parts of the FS and hence leaves a sufficiently large portion of the FS for the SC.
The reduction of the DOS by the partial FS gapping can explain d$T_c$/d$P$ $<$ 0 for $P$ $\sim$ 8--10 kbar near the onset of AFM order.
As the growth of the DOS suppression with pressure slows down, $T_c$ increases again probably because of increasing spin fluctuations.

The occurrence of the FS reconstruction at the onset of the AFM order would suggest that the ordering wave vector is likely ($\pi$, 0) (or very close to it): otherwise the hole and electron FS cylinders would not intersect each other to open a gap, as they are very small.
This would raise an interesting question about the crystal structure for $P$ $>$ $\sim$18 kbar where the structural transition $T_s$ is no longer observed (Fig.~\ref{phase}). 
Since a single-$q$ order with ($\pi$, 0) is incompatible with the tetragonal structure, the AFM transition at $T_N$ at those pressures might be accompanied by an orthorhombic structural distortion similar to the AFM transition in BaFe$_2$As$_2$ \cite{Rotter08PRB}.
An alternative scenario would be a double-$q$ order preserving the tetragonal symmetry.
In addition, theoretical studies based on \textit{ab initio} calculations suggest that the absence of magnetic order at ambient pressure is due to competition between different magnetic ordering vectors and that the application of pressure lifts this near-degeneracy, leading to a ($\pi$, 0) stripe order \cite{Glasbrenner15NatPhys} or a ($\pi$, $\pi$/2) staggered dimer order \cite{Hirayama15JPSJ}.
To settle those issues, neutron diffraction measurements under high pressure are highly desirable.

Recently, Kaluarachchi \textit{et al.} have reported a high-pressure study of the upper critical field in FeSe \cite{Kaluarachchi16PRB}.
They have found abrupt changes in the normalized slope of the $c$-axis upper critical field $(-1/T_c)(\mathrm{d}B_{c2, c}/\mathrm{d}T)|_{T_c}$ at $P_1 \sim$8 kbar, the onset of the AFM order, and $P_2 \sim$12 kbar, the local minimum of $T_c$, and have shown that the observed slopes in the three regions, i.e., $P < P_1$, $P_1 < P <P_2$, and $P_2 < P$, can be explained very well by the Fermi surface parameters determined in the present study (Fig.~\ref{param}).
They have therefore concluded that the abrupt changes in the slope are due to changes of the Fermi surface.
The FS change at $P_1$ agrees with our proposed AFM reconstruction of the FS.
The second one at $P_2$ is compatible with the discontinuous change of the frequency from $\Omega$ to $\Psi$ across $P_2$.
The $\mu$SR measurements suggest that the AFM order changes from incommensurate to commensurate at $P_2$ \cite{Bendele10PRL, Bendele12PRB}, which could be related to this FS change.

Finally, we note that the ratio $k_BT_c/E_F$ remains as high as $\sim$ 0.1 even at $P$ = 16.1 kbar [Fig.~\ref{param}(d)].
The smallness of $E_F$ (relative to $T_c$) seems to be a key factor to be kept in mind when we tackle not only the peculiar SC in FeSe at ambient pressure but also the huge enhancement of $T_c$ under high pressure.

\section{Summary}

We have performed SdH oscillation measurements on FeSe under high pressure up to $P$ = 16.1 kbar and found a sudden change in the SdH oscillations at the onset of the AFM order.
We have shown that the pressure inhomogeneity or phase separation is unlikely to explain the observed change and argued that it is due to the FS reconstruction by the AFM order.
The negative d$T_c$/d$P$ at the onset of the AFM may be attributed to the reduction in the density of states due to the reconstruction.
The AFM ordering vector could be ($\pi$, 0) and further studies on this point are necessary. 
The ratio $k_BT_c/E_F$ remains high in the investigated pressure range.

\begin{acknowledgments}
This work has been supported by a Grant-in-Aid for Scientific Research on Innovative Areas ``Topological Materials Science'' (KAKENHI Grant No. 15H05852), Japan-Germany Research Cooperative Program, KAKENHI from JSPS and Project No. 56393598 from DAAD, and JSPS KAKENHI Grant Number 26400373.
A portion of this work was performed at the National High Magnetic Field Laboratory, which is supported by National Science Foundation Cooperative Agreement No. DMR-1157490 and the State of Florida.
TT thanks Ola Kenji Forslund for technical assistance.
\end{acknowledgments}

\appendix

\section{Maximum entropy method (MEM) of spectral analysis and quantum oscillation studies}

\subsection{Introduction}

We can observe physical phenomena only in a limited window, i.e., in a limited period of time ( or field in the present case).
When estimating the power spectrum of the observed physical quantity, we have to assume how it behaves outside the experimental window.

The Fourier transform (FT) analysis assumes that the observed behavior is repeated infinitely outside the window.
Therefore, the frequency resolution $\Delta F$ is basically limited by the reciprocal data length (though it is usually ``enhanced'' by zero padding): in the case of quantum oscillation measurements, $\Delta F =[\Delta(1/B)]^{-1}$, where $\Delta (1/B)$ is the width of the measured field range in $1/B$.

On the other hand, the MEM fits an autoregressive model to the observed behavior and assumes that the physical quantity evolves outside the window according to this model.
In a sense, it analyzes infinitely long data.
Because of this, it can produce fine spectral peaks even for short data.

The MEM has been used in the fields of geophysics and astrophysics since late 1960¡Çs \cite{Ulrych75RGSP}.
In those research areas, the time scale of physical phenomena does not always allow one to observe many oscillation periods, and hence one has to work with short data.
The MEM has proven to be a powerful tool to analyze short data in those areas.

In the area of quantum oscillation studies, Sigfusson \textit{et al}. reported application of MEM to spectral analysis of de Haas-van Alphen oscillations in 1992 \cite{Sigfusson92PRB}.
Since then MEM has been used occasionally because of its superior frequency resolution \cite{Nimori93PhysB, Harrison93JPCM, Uji95SynthM, Terashima98JMMM, Coldea04PRB, Kang09PRB}.
However, it has scarcely been used to determine effective masses, which requires a reliable estimation of oscillation amplitudes.
The limited use of MEM is probably due to the fact that peak values of MEM PSD are so sensitive to noise in analyzed data that they can not be used as a measure of oscillation amplitudes.
This obstacle can be overcome by using integrated power \cite{Ulrych75RGSP}.
In the following, after a brief description of the MEM, we demonstrate this approach by using simulated data.
Then, we consider precautions to be taken when applying MEM to effective-mass analysis and finally show a comparison between MEM and FT analyses of the present data.
	
\subsection{MEM spectral analysis}

MEM spectral analysis fits an autoregressive (AR) model to experimental data and estimates its power spectrum.
The AR model of order $M$ fitted to a time series of length $N$ ($M < N$) and interval $\Delta t$ is expressed as \cite{Press02}:
\begin{equation}
x_n = -\sum_{m=1}^M a_{M, m}x_{n-m} + \epsilon_n,
\end{equation}
where $x_n$ is the $n$th data point and $\epsilon_n$ is a white noise with a variance $\sigma^2$.
The MEM PSD is given by:
\begin{equation}
S(f) = \frac{\sigma^2 \Delta t}{|1 + \sum_{m=1}^M a_{M, m} \exp(-2\pi ifm \Delta t)|^2}
\end{equation}
(in analyses of quantum oscillations $1/B$ corresponds to time).

The AR coefficients $a_{M, m}$ are estimated by using Burg's or other methods \cite{Press02, Ulrych75RGSP}.
The order $M$ is determined empirically by comparing spectra with those obtained by FT, for example:
With increasing $M$, initially the number of spectral peaks increases.
Then, the peak number stays nearly constant for a fairly long interval of $M$ and the peaks sharpen.
Finally, the peaks start splitting and proliferate.
$M$ should be chosen in the stable region, and the spectrum should match that by FT reasonably well.
It is often suggested that approximately $N/2$ is a safe maximum for $M$ \cite{Ulrych75RGSP, Pardo05ComputGeosci}.

\subsection{Amplitude estimation by MEM}

We analyze signals of the form:
\begin{equation}
A_1\sin(2\pi F_1/B-3\pi /4)+A_2\sin(2\pi F_2/B-5\pi /4)+\mathrm{(noise)},
\end{equation}
where $F_1$ = 90 T, $F_2$ = 195 T, and 28 T $<$ $B$ $<$ 45 T.
For each of ($A_1$, $A_2$) = (1, 0.7), (0.8, 0.5), and (0.6, 0.3), we generate three signals by adding random noise with a Gaussian distribution of a standard deviation of 0.04.
The number of data points is 1024, evenly spaced in $1/B$, and the order of the AR model is $M$ = 400.
Note that the frequency $F_1$ has only 1.2 periods in the generated field range.

Our MEM code is based on \cite{Press02}.
However, since the power scaling used in \cite{Press02} is inconvenient for the present purpose, PSDs are multiplied by 4$\Delta$, where $\Delta$ is the sampling interval [i.e., $\Delta(1/B)$ between two adjacent data points in the present case].
This gives the correct amplitude $A$ = 1 for $\sin(x)$.

\begin{figure}
\includegraphics[width=8.cm]{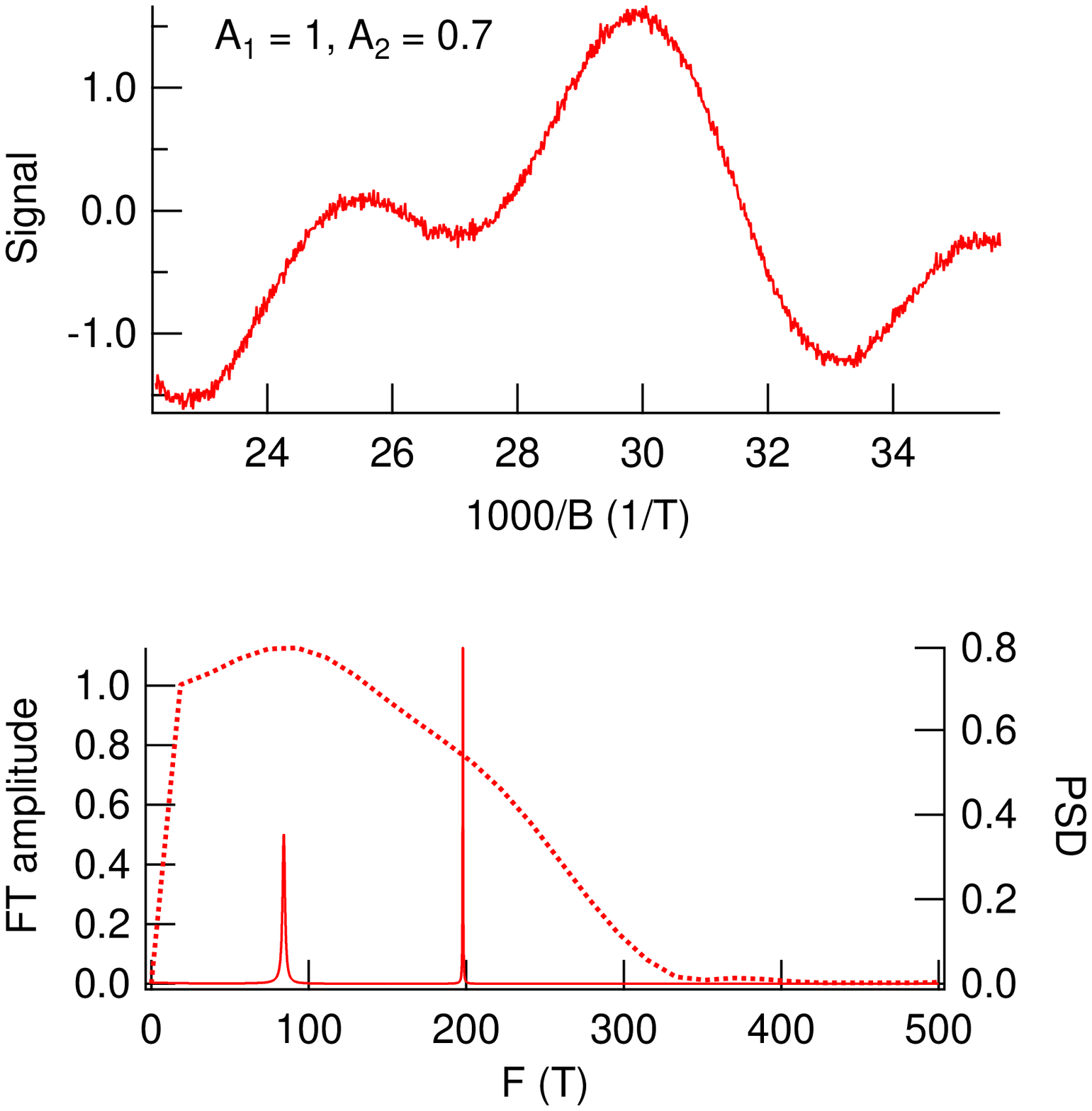}
\caption{\label{sig10}(Color online).  (Upper) Example of simulated signal.  (Lower) FT amplitude (dotted curve) and MEM PSD (solid curve).}   
\end{figure}

\begin{figure}
\includegraphics[width=8.cm]{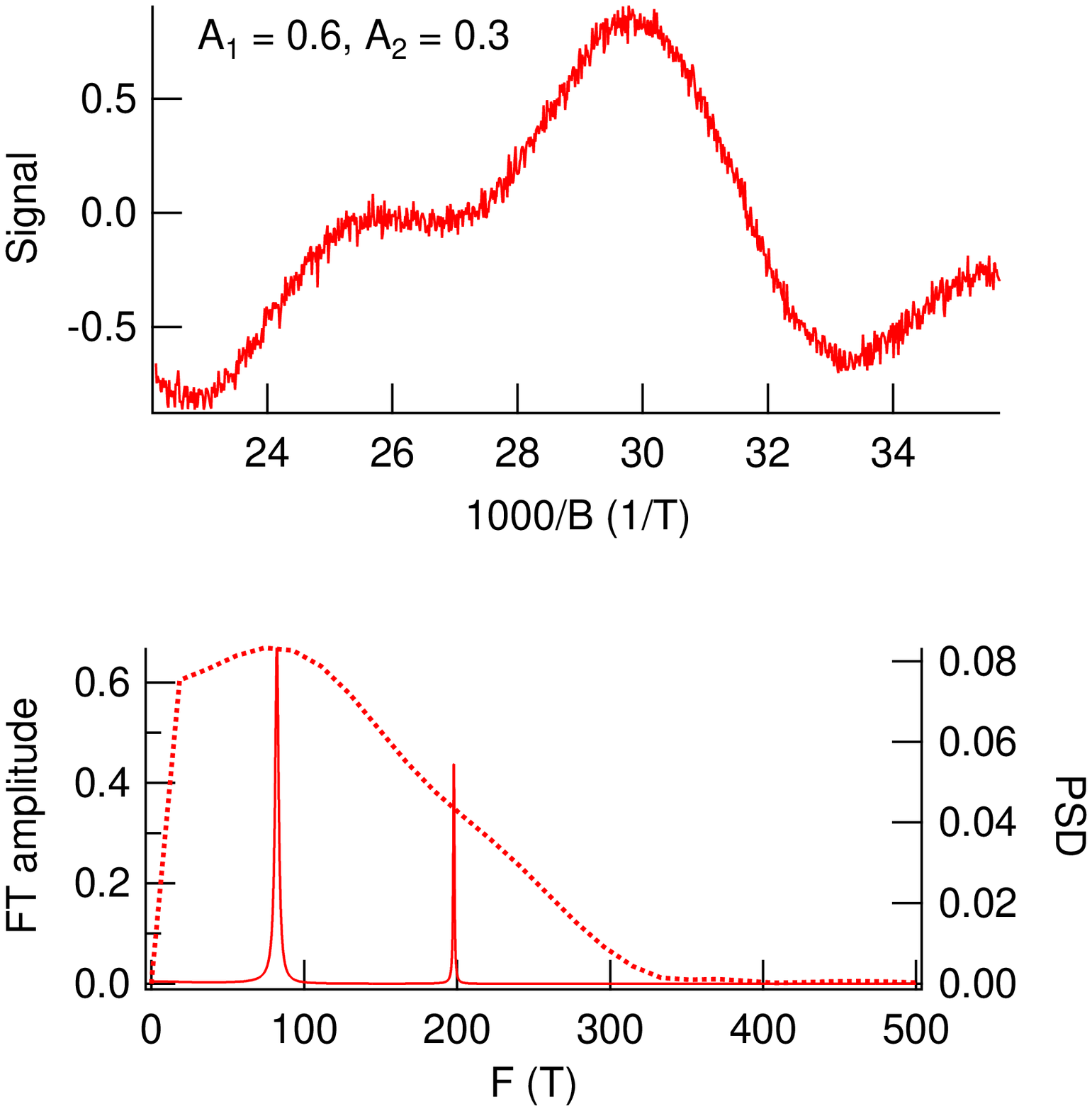}
\caption{\label{sig18}(Color online).  (Upper) Example of simulated signal.  (Lower) FT amplitude (dotted curve) and MEM PSD (solid curve).}   
\end{figure}

\begin{figure}
\includegraphics[width=8.cm]{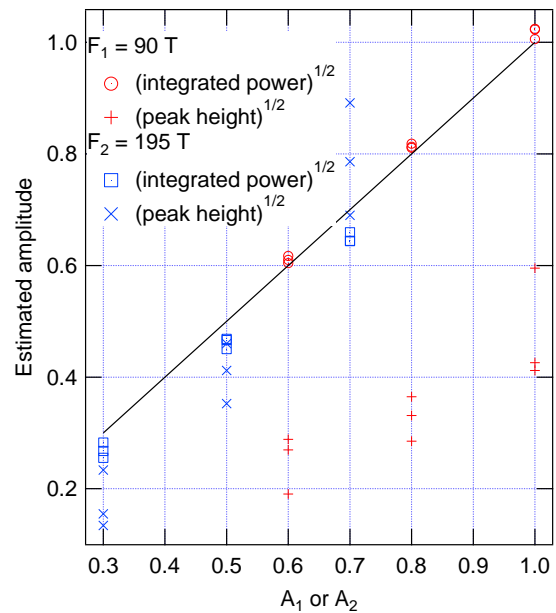}
\caption{\label{Amp}(Color online).  Estimated amplitude vs. $A_1$ and $A_2$.  Circles and squares are square roots of integrated powers, while + and x marks are square roots of peak heights.}   
\end{figure}

Examples of simulated signals, FT amplitudes, and MEM PSDs are shown in Figs.~\ref{sig10} and \ref{sig18}.
While FT can not resolve the two frequencies because the field range is too narrow, MEM can.
MEM gives $F_1$ = 84(2) T and $F_2$ = 199(1) T, which are within 7 and 2\% of the true values, respectively. 
However, it is clear that heights of MEM PSD peaks are not reliable: in Fig.~\ref{sig10}, the height of the $F_2$ peak is more than two times larger than that of $F_1$ despite $A_1 > A_2$.

Figure~\ref{Amp} plots square roots of the peak heights against $A_1$ and $A_2$ (denoted by + and x, respectively), where a large scatter of the data points is also apparent.
(Power is the squared oscillation amplitude.)
On the other hand, the square root of the integrated power under a peak gives an accurate measure of the amplitude (circles and squares).
Here we have estimated the integrated power from the peak height and width assuming a Lorentzian peak shape, which is generally a very good approximation.
For each value of $A_1$ or $A_2$, the three estimates converge satisfactorily.
On average, deviations from the true amplitudes are only 2 and -8\% for $A_1$ and $A_2$. respectively.
To summarize, the square root of the integrated power in MEM PSD can safely be used as a measure of oscillation amplitudes.

\subsection{Practical considerations in application of MEM to effective-mass analysis}

FT analysis is usually performed with data windowing, whereas MEM is used without it.
This seems to necessitate special precautions when applying MEM to effective-mass analysis.
A window function takes a maximum at the center of the window and smoothly goes to zero at both ends.
Windowing enhances contributions from the center of the window to obtain an amplitude estimate. 
Hence the estimated FT amplitude can usually be assumed to be a good approximation to the oscillation amplitude at the center of the window (i.e., the center of the field range in $1/B$ in the present case).
However, each region of the window contributes equally to MEM PSD.
If oscillations grow significantly with field, the estimated amplitude is biased by the high-field region and hence may not be a good approximation to the amplitude at the center of the window.
It is therefore better to use a narrow field range.
Secondly, when a background is subtracted, residuals may be large near an end of the fitting region.
This is not a problem in FT analysis because those residuals are suppressed by windowing, but it is in MEM analysis.
It is better to omit a region where residuals are large in MEM analysis.

In addition, the following caveat is in order.
MEM gives reliable frequency estimates even for a very narrow data window as shown above: the $F_1$ oscillation has only 1.2 periods in the window. 
However, such short data can in reality not be used for the mass analysis because the amplitude of extremely slow oscillations can crucially depend on the background subtraction: sometimes the amplitude can be suppressed by the background subtraction very effectively.
Our criterion is to have at least two oscillation periods in a data window so that the oscillation will not inadvertently be eliminated by subtraction of a low-order polynomial.

\subsection{Example of application of MEM to effective-mass analysis}

\begin{table*}
\caption{\label{Tab} Frequencies and effective masses estimated by FT and MEM.}
\begin{ruledtabular}
\begin{tabular}{ccccccc}
& \multicolumn{2}{c}{FT ($B$ = 26 - 45 T)} & \multicolumn{2}{c}{MEM ($B$ = 27.7 - 45 T)} & \multicolumn{2}{c}{MEM ($B$ = 34.2 - 45 T)}\\
\cline{2-3} \cline{4-5} \cline{6-7}
Branch & $F$ (kT) & $m^*/m_e$ & $F$ (kT) & $m^*/m_e$ & $F$ (kT) & $m^*/m_e$ \\
\hline
2$\alpha$ &  &  & 0.139(9) & 4.4(3) \\
$\beta$ & 0.256(6) & 4.1(2) & 0.24(2) & 4.3(2) \\
$\gamma$ & 0.740(5) & 7.6(3) & 0.736(2) & 5.8(4) & 0.754(9) & 6.4(3) \\
$\delta$ & 0.84(2) & 5.2(2) & 0.847(3) & 4.4(3) & 0.873(7) & 4.3(3)\\
\end{tabular}
\end{ruledtabular}
\end{table*}

\begin{figure}[t]
\includegraphics[width=8.6cm]{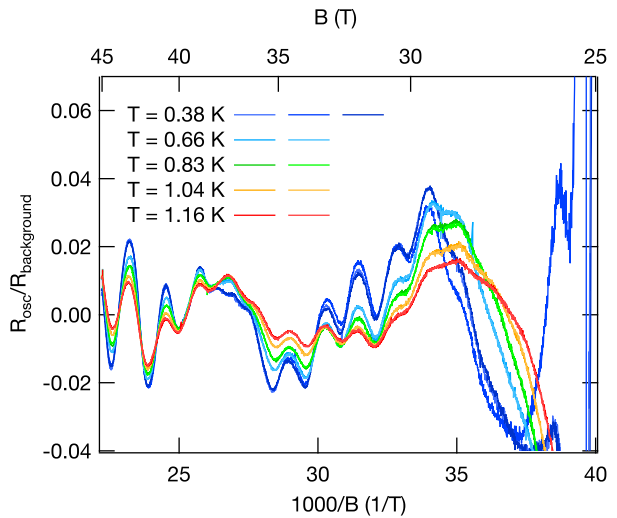}
\caption{\label{Sig4kbar}(Color online).  SdH oscillations in FeSe at $P$ = 4.4 kbar.}   
\end{figure}

\begin{figure}[t]
\includegraphics[width=8.6cm]{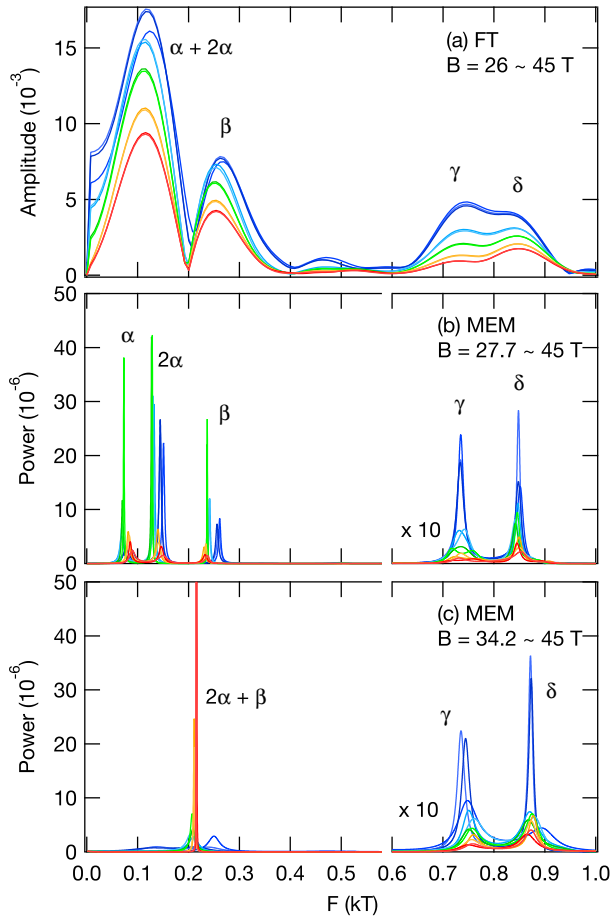}
\caption{\label{FTMEM4kbar}(Color online).  (a) FT amplitudes and (b, c) MEM PSDs.}   
\end{figure}

\begin{figure*}[t]
\includegraphics[width=17cm]{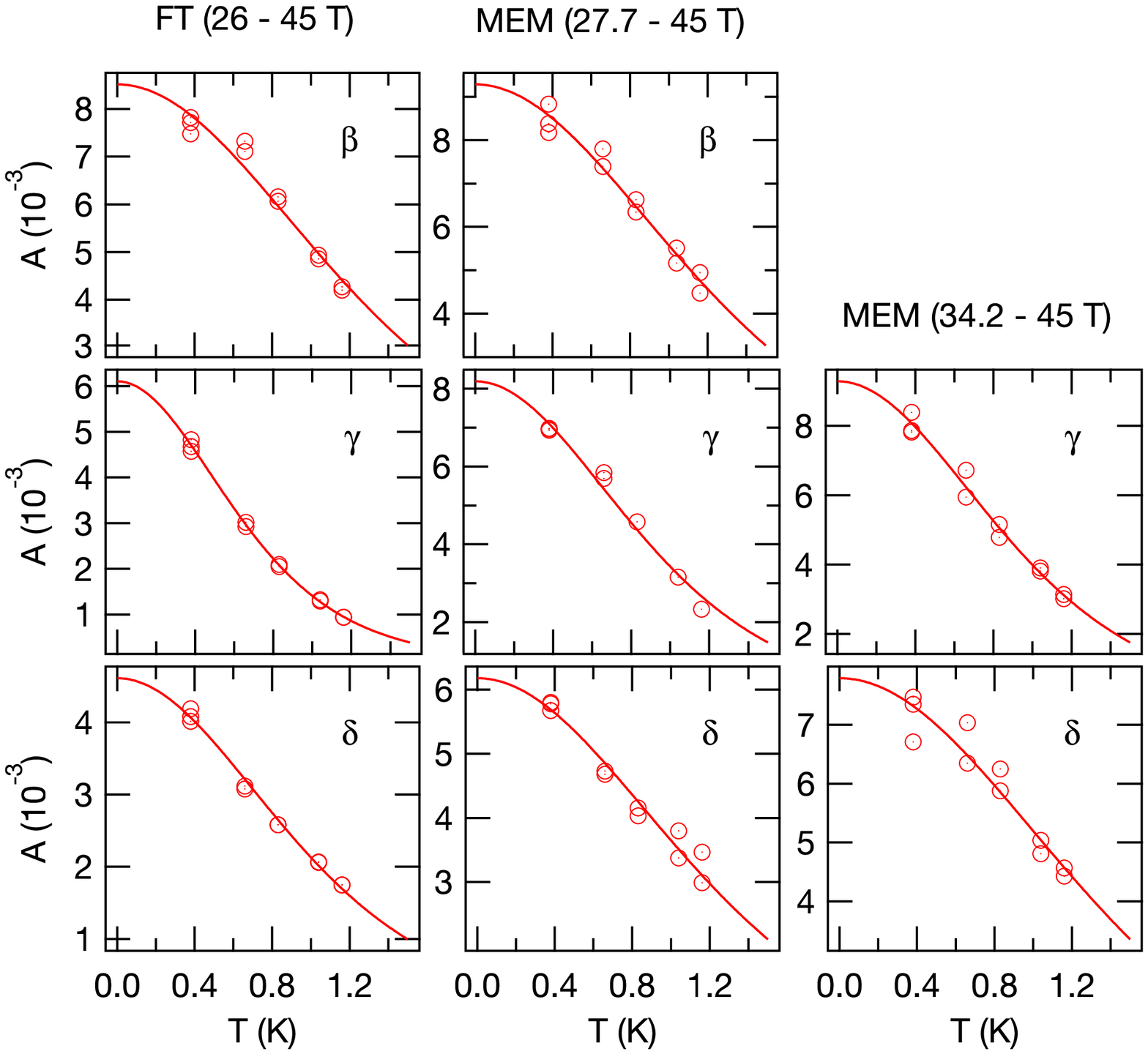}
\caption{\label{mass4kbar}(Color online).  Temperature dependence of oscillation amplitudes estimated by FT and MEM.  Solid curves are fits to the Lifshitz-Kosevich formula to estimate effective masses \cite{Shoenberg84}.}   
\end{figure*}

Finally, we compare FT and MEM analyses of SdH oscillations in FeSe at $P$ = 4.4 kbar.
Figure~\ref{Sig4kbar} shows oscillations measured at different temperatures.
Two field sweeps were made at each temperature (three at $T$ = 0.38 K).
Figure~\ref{FTMEM4kbar} shows FT and MEM spectra.
FT was applied in a field range between $B$ = 26 and 45 T.
MEM was applied in a slightly narrower range $B$ = 27.7 -- 45 T to avoid large residuals of background subtraction near $B$ =26 T.
The data in this field range were interpolated to $N$ = 1024 points, which are equally spaced in 1/$B$, and then PSDs were calculated with $M$ = 650 [Fig.~\ref{FTMEM4kbar}(b)].
Also, PSDs for the high-field half of the same data ($B$ = 34.2 -- 45 T, $N$ = 512) were calculated with $M$ = 300 [Fig.~\ref{FTMEM4kbar}(c)].

The FT spectra can resolve $\beta$, $\gamma$, and $\delta$, while the MEM spectra for $B$ = 27.7 -- 45 T can resolve $\alpha$ and 2$\alpha$ as well.
The MEM spectra for $B$ = 34.2 -- 45 T resolve only $\gamma$ and $\delta$ because of the narrow field range.

Figure~\ref{mass4kbar} shows temperature dependences of amplitudes of those resolved frequencies (except $\alpha$ and 2$\alpha$).
For the MEM spectra, integrated powers were estimated by fitting Lorentzian functions to spectral peaks.
The solid curves are fits to the Lifshitz-Kosevich formula to estimate effective masses \cite{Shoenberg84}.

Table~\ref{Tab} shows frequencies and effective masses determined in those analyses.
Errors for $F$ are standard deviations of frequencies determined in different measurements, and those for $m^*$ are fitting uncertainties.
The $\alpha$ frequency has only one oscillation period in the range $B$ = 27.7 -- 45 T, and hence its temperature dependence was not analyzed.
Instead, the temperature dependence of the 2$\alpha$ oscillation was analyzed, which is expected to give a two times larger effective mass than that of $\alpha$.
The frequencies determined by FT and MEM (for both field windows) almost agree within the errors.
Comparing the FT and wide-region MEM ($B$ = 27.7 -- 45 T) analyses, the effective mass for $\beta$ agrees well, while those for $\gamma$ and $\delta$ do not.
Using the narrow field range ($B$ = 34.2 -- 45 T) for MEM, the agreement with the FT analysis is improved for $\gamma$, though virtually no change is seen for $\delta$.
As discussed above, effective-mass estimates by MEM can be more prone to be influenced by the field dependence of oscillation amplitudes, and estimates of heavy masses can be more biased because of stronger field dependences of amplitudes.
Thus the narrow-range MEM results seem more reliable, allthough the masses for $\gamma$ and $\delta$ differ beyond the fitting errors even between the FT and narrow-region MEM analyses.


\begin{thebibliography}{67}%
\makeatletter
\providecommand \@ifxundefined [1]{%
 \@ifx{#1\undefined}
}%
\providecommand \@ifnum [1]{%
 \ifnum #1\expandafter \@firstoftwo
 \else \expandafter \@secondoftwo
 \fi
}%
\providecommand \@ifx [1]{%
 \ifx #1\expandafter \@firstoftwo
 \else \expandafter \@secondoftwo
 \fi
}%
\providecommand \natexlab [1]{#1}%
\providecommand \enquote  [1]{``#1''}%
\providecommand \bibnamefont  [1]{#1}%
\providecommand \bibfnamefont [1]{#1}%
\providecommand \citenamefont [1]{#1}%
\providecommand \href@noop [0]{\@secondoftwo}%
\providecommand \href [0]{\begingroup \@sanitize@url \@href}%
\providecommand \@href[1]{\@@startlink{#1}\@@href}%
\providecommand \@@href[1]{\endgroup#1\@@endlink}%
\providecommand \@sanitize@url [0]{\catcode `\\12\catcode `\$12\catcode
  `\&12\catcode `\#12\catcode `\^12\catcode `\_12\catcode `\%12\relax}%
\providecommand \@@startlink[1]{}%
\providecommand \@@endlink[0]{}%
\providecommand \url  [0]{\begingroup\@sanitize@url \@url }%
\providecommand \@url [1]{\endgroup\@href {#1}{\urlprefix }}%
\providecommand \urlprefix  [0]{URL }%
\providecommand \Eprint [0]{\href }%
\providecommand \doibase [0]{http://dx.doi.org/}%
\providecommand \selectlanguage [0]{\@gobble}%
\providecommand \bibinfo  [0]{\@secondoftwo}%
\providecommand \bibfield  [0]{\@secondoftwo}%
\providecommand \translation [1]{[#1]}%
\providecommand \BibitemOpen [0]{}%
\providecommand \bibitemStop [0]{}%
\providecommand \bibitemNoStop [0]{.\EOS\space}%
\providecommand \EOS [0]{\spacefactor3000\relax}%
\providecommand \BibitemShut  [1]{\csname bibitem#1\endcsname}%
\let\auto@bib@innerbib\@empty
\bibitem [{\citenamefont {Kamihara}\ \emph {et~al.}(2008)\citenamefont
  {Kamihara}, \citenamefont {Watanabe}, \citenamefont {Hirano},\ and\
  \citenamefont {Hosono}}]{Kamihara08JACS}%
  \BibitemOpen
  \bibfield  {author} {\bibinfo {author} {\bibfnamefont {Y.}~\bibnamefont
  {Kamihara}}, \bibinfo {author} {\bibfnamefont {T.}~\bibnamefont {Watanabe}},
  \bibinfo {author} {\bibfnamefont {M.}~\bibnamefont {Hirano}}, \ and\ \bibinfo
  {author} {\bibfnamefont {H.}~\bibnamefont {Hosono}},\ }\href {\doibase
  10.1021/ja800073m} {\bibfield  {journal} {\bibinfo  {journal} {J. Am. Chem.
  Soc.}\ }\textbf {\bibinfo {volume} {130}},\ \bibinfo {pages} {3296} (\bibinfo
  {year} {2008})}\BibitemShut {NoStop}%
\bibitem [{\citenamefont {Fernandes}\ \emph {et~al.}(2014)\citenamefont
  {Fernandes}, \citenamefont {Chubukov},\ and\ \citenamefont
  {Schmalian}}]{Fernandes14NatPhys}%
  \BibitemOpen
  \bibfield  {author} {\bibinfo {author} {\bibfnamefont {R.~M.}\ \bibnamefont
  {Fernandes}}, \bibinfo {author} {\bibfnamefont {A.~V.}\ \bibnamefont
  {Chubukov}}, \ and\ \bibinfo {author} {\bibfnamefont {J.}~\bibnamefont
  {Schmalian}},\ }\href {http://dx.doi.org/10.1038/nphys2877} {\bibfield
  {journal} {\bibinfo  {journal} {Nat. Phys.}\ }\textbf {\bibinfo {volume}
  {10}},\ \bibinfo {pages} {97} (\bibinfo {year} {2014})}\BibitemShut {NoStop}%
\bibitem [{\citenamefont {Rotter}\ \emph
  {et~al.}(2008{\natexlab{a}})\citenamefont {Rotter}, \citenamefont {Tegel},\
  and\ \citenamefont {Johrendt}}]{Rotter08PRL}%
  \BibitemOpen
  \bibfield  {author} {\bibinfo {author} {\bibfnamefont {M.}~\bibnamefont
  {Rotter}}, \bibinfo {author} {\bibfnamefont {M.}~\bibnamefont {Tegel}}, \
  and\ \bibinfo {author} {\bibfnamefont {D.}~\bibnamefont {Johrendt}},\ }\href
  {\doibase 10.1103/PhysRevLett.101.107006} {\bibfield  {journal} {\bibinfo
  {journal} {Phys. Rev. Lett.}\ }\textbf {\bibinfo {volume} {101}},\ \bibinfo
  {eid} {107006} (\bibinfo {year} {2008}{\natexlab{a}})}\BibitemShut {NoStop}%
\bibitem [{\citenamefont {Sasmal}\ \emph {et~al.}(2008)\citenamefont {Sasmal},
  \citenamefont {Lv}, \citenamefont {Lorenz}, \citenamefont {Guloy},
  \citenamefont {Chen}, \citenamefont {Xue},\ and\ \citenamefont
  {Chu}}]{Sasmal08PRL}%
  \BibitemOpen
  \bibfield  {author} {\bibinfo {author} {\bibfnamefont {K.}~\bibnamefont
  {Sasmal}}, \bibinfo {author} {\bibfnamefont {B.}~\bibnamefont {Lv}}, \bibinfo
  {author} {\bibfnamefont {B.}~\bibnamefont {Lorenz}}, \bibinfo {author}
  {\bibfnamefont {A.~M.}\ \bibnamefont {Guloy}}, \bibinfo {author}
  {\bibfnamefont {F.}~\bibnamefont {Chen}}, \bibinfo {author} {\bibfnamefont
  {Y.-Y.}\ \bibnamefont {Xue}}, \ and\ \bibinfo {author} {\bibfnamefont
  {C.-W.}\ \bibnamefont {Chu}},\ }\href {\doibase
  10.1103/PhysRevLett.101.107007} {\bibfield  {journal} {\bibinfo  {journal}
  {Phys. Rev. Lett.}\ }\textbf {\bibinfo {volume} {101}},\ \bibinfo {eid}
  {107007} (\bibinfo {year} {2008})}\BibitemShut {NoStop}%
\bibitem [{\citenamefont {Hsu}\ \emph {et~al.}(2008)\citenamefont {Hsu},
  \citenamefont {Luo}, \citenamefont {Yeh}, \citenamefont {Chen}, \citenamefont
  {Huang}, \citenamefont {Wu}, \citenamefont {Lee}, \citenamefont {Huang},
  \citenamefont {Chu}, \citenamefont {Yan},\ and\ \citenamefont
  {Wu}}]{Hsu08PNAS}%
  \BibitemOpen
  \bibfield  {author} {\bibinfo {author} {\bibfnamefont {F.-C.}\ \bibnamefont
  {Hsu}}, \bibinfo {author} {\bibfnamefont {J.-Y.}\ \bibnamefont {Luo}},
  \bibinfo {author} {\bibfnamefont {K.-W.}\ \bibnamefont {Yeh}}, \bibinfo
  {author} {\bibfnamefont {T.-K.}\ \bibnamefont {Chen}}, \bibinfo {author}
  {\bibfnamefont {T.-W.}\ \bibnamefont {Huang}}, \bibinfo {author}
  {\bibfnamefont {P.~M.}\ \bibnamefont {Wu}}, \bibinfo {author} {\bibfnamefont
  {Y.-C.}\ \bibnamefont {Lee}}, \bibinfo {author} {\bibfnamefont {Y.-L.}\
  \bibnamefont {Huang}}, \bibinfo {author} {\bibfnamefont {Y.-Y.}\ \bibnamefont
  {Chu}}, \bibinfo {author} {\bibfnamefont {D.-C.}\ \bibnamefont {Yan}}, \ and\
  \bibinfo {author} {\bibfnamefont {M.-K.}\ \bibnamefont {Wu}},\ }\href
  {\doibase 10.1073/pnas.0807325105} {\bibfield  {journal} {\bibinfo  {journal}
  {Proc. Nat. Acad. Sci. U. S. A.}\ }\textbf {\bibinfo {volume} {105}},\
  \bibinfo {pages} {14262} (\bibinfo {year} {2008})}\BibitemShut {NoStop}%
\bibitem [{\citenamefont {Mizuguchi}\ \emph {et~al.}(2008)\citenamefont
  {Mizuguchi}, \citenamefont {Tomioka}, \citenamefont {Tsuda}, \citenamefont
  {Yamaguchi},\ and\ \citenamefont {Takano}}]{Mizuguchi08APL}%
  \BibitemOpen
  \bibfield  {author} {\bibinfo {author} {\bibfnamefont {Y.}~\bibnamefont
  {Mizuguchi}}, \bibinfo {author} {\bibfnamefont {F.}~\bibnamefont {Tomioka}},
  \bibinfo {author} {\bibfnamefont {S.}~\bibnamefont {Tsuda}}, \bibinfo
  {author} {\bibfnamefont {T.}~\bibnamefont {Yamaguchi}}, \ and\ \bibinfo
  {author} {\bibfnamefont {Y.}~\bibnamefont {Takano}},\ }\href {\doibase
  http://dx.doi.org/10.1063/1.3000616} {\bibfield  {journal} {\bibinfo
  {journal} {Appl. Phys. Lett.}\ }\textbf {\bibinfo {volume} {93}},\ \bibinfo
  {eid} {152505} (\bibinfo {year} {2008})}\BibitemShut {NoStop}%
\bibitem [{\citenamefont {Medvedev}\ \emph {et~al.}(2009)\citenamefont
  {Medvedev}, \citenamefont {McQueen}, \citenamefont {Troyan}, \citenamefont
  {Palasyuk}, \citenamefont {Eremets}, \citenamefont {Cava}, \citenamefont
  {Naghavi}, \citenamefont {Casper}, \citenamefont {Ksenofontov}, \citenamefont
  {Wortmann},\ and\ \citenamefont {Felser}}]{Medvedev09Nmat}%
  \BibitemOpen
  \bibfield  {author} {\bibinfo {author} {\bibfnamefont {S.}~\bibnamefont
  {Medvedev}}, \bibinfo {author} {\bibfnamefont {T.~M.}\ \bibnamefont
  {McQueen}}, \bibinfo {author} {\bibfnamefont {I.~A.}\ \bibnamefont {Troyan}},
  \bibinfo {author} {\bibfnamefont {T.}~\bibnamefont {Palasyuk}}, \bibinfo
  {author} {\bibfnamefont {M.~I.}\ \bibnamefont {Eremets}}, \bibinfo {author}
  {\bibfnamefont {R.~J.}\ \bibnamefont {Cava}}, \bibinfo {author}
  {\bibfnamefont {S.}~\bibnamefont {Naghavi}}, \bibinfo {author} {\bibfnamefont
  {F.}~\bibnamefont {Casper}}, \bibinfo {author} {\bibfnamefont
  {V.}~\bibnamefont {Ksenofontov}}, \bibinfo {author} {\bibfnamefont
  {G.}~\bibnamefont {Wortmann}}, \ and\ \bibinfo {author} {\bibfnamefont
  {C.}~\bibnamefont {Felser}},\ }\href@noop {} {\bibfield  {journal} {\bibinfo
  {journal} {Nat. Mater.}\ }\textbf {\bibinfo {volume} {8}},\ \bibinfo {pages}
  {630} (\bibinfo {year} {2009})}\BibitemShut {NoStop}%
\bibitem [{\citenamefont {Wang}\ \emph {et~al.}(2012)\citenamefont {Wang},
  \citenamefont {Li}, \citenamefont {Zhang}, \citenamefont {Zhang},
  \citenamefont {Zhang}, \citenamefont {Li}, \citenamefont {Ding},
  \citenamefont {Ou}, \citenamefont {Deng}, \citenamefont {Chang},
  \citenamefont {Wen}, \citenamefont {Song}, \citenamefont {He}, \citenamefont
  {Jia}, \citenamefont {Ji}, \citenamefont {Wang}, \citenamefont {Wang},
  \citenamefont {Chen}, \citenamefont {Ma},\ and\ \citenamefont
  {Xue}}]{Wang12CPL}%
  \BibitemOpen
  \bibfield  {author} {\bibinfo {author} {\bibfnamefont {Q.-Y.}\ \bibnamefont
  {Wang}}, \bibinfo {author} {\bibfnamefont {Z.}~\bibnamefont {Li}}, \bibinfo
  {author} {\bibfnamefont {W.-H.}\ \bibnamefont {Zhang}}, \bibinfo {author}
  {\bibfnamefont {Z.-C.}\ \bibnamefont {Zhang}}, \bibinfo {author}
  {\bibfnamefont {J.-S.}\ \bibnamefont {Zhang}}, \bibinfo {author}
  {\bibfnamefont {W.}~\bibnamefont {Li}}, \bibinfo {author} {\bibfnamefont
  {H.}~\bibnamefont {Ding}}, \bibinfo {author} {\bibfnamefont {Y.-B.}\
  \bibnamefont {Ou}}, \bibinfo {author} {\bibfnamefont {P.}~\bibnamefont
  {Deng}}, \bibinfo {author} {\bibfnamefont {K.}~\bibnamefont {Chang}},
  \bibinfo {author} {\bibfnamefont {J.}~\bibnamefont {Wen}}, \bibinfo {author}
  {\bibfnamefont {C.-L.}\ \bibnamefont {Song}}, \bibinfo {author}
  {\bibfnamefont {K.}~\bibnamefont {He}}, \bibinfo {author} {\bibfnamefont
  {J.-F.}\ \bibnamefont {Jia}}, \bibinfo {author} {\bibfnamefont {S.-H.}\
  \bibnamefont {Ji}}, \bibinfo {author} {\bibfnamefont {Y.-Y.}\ \bibnamefont
  {Wang}}, \bibinfo {author} {\bibfnamefont {L.-L.}\ \bibnamefont {Wang}},
  \bibinfo {author} {\bibfnamefont {X.}~\bibnamefont {Chen}}, \bibinfo {author}
  {\bibfnamefont {X.-C.}\ \bibnamefont {Ma}}, \ and\ \bibinfo {author}
  {\bibfnamefont {Q.-K.}\ \bibnamefont {Xue}},\ }\href
  {http://stacks.iop.org/0256-307X/29/i=3/a=037402} {\bibfield  {journal}
  {\bibinfo  {journal} {Chin. Phys. Lett.}\ }\textbf {\bibinfo {volume} {29}},\
  \bibinfo {pages} {037402} (\bibinfo {year} {2012})}\BibitemShut {NoStop}%
\bibitem [{\citenamefont {McQueen}\ \emph {et~al.}(2009)\citenamefont
  {McQueen}, \citenamefont {Williams}, \citenamefont {Stephens}, \citenamefont
  {Tao}, \citenamefont {Zhu}, \citenamefont {Ksenofontov}, \citenamefont
  {Casper}, \citenamefont {Felser},\ and\ \citenamefont {Cava}}]{McQueen09PRL}%
  \BibitemOpen
  \bibfield  {author} {\bibinfo {author} {\bibfnamefont {T.~M.}\ \bibnamefont
  {McQueen}}, \bibinfo {author} {\bibfnamefont {A.~J.}\ \bibnamefont
  {Williams}}, \bibinfo {author} {\bibfnamefont {P.~W.}\ \bibnamefont
  {Stephens}}, \bibinfo {author} {\bibfnamefont {J.}~\bibnamefont {Tao}},
  \bibinfo {author} {\bibfnamefont {Y.}~\bibnamefont {Zhu}}, \bibinfo {author}
  {\bibfnamefont {V.}~\bibnamefont {Ksenofontov}}, \bibinfo {author}
  {\bibfnamefont {F.}~\bibnamefont {Casper}}, \bibinfo {author} {\bibfnamefont
  {C.}~\bibnamefont {Felser}}, \ and\ \bibinfo {author} {\bibfnamefont {R.~J.}\
  \bibnamefont {Cava}},\ }\href {\doibase 10.1103/PhysRevLett.103.057002}
  {\bibfield  {journal} {\bibinfo  {journal} {Phys. Rev. Lett.}\ }\textbf
  {\bibinfo {volume} {103}},\ \bibinfo {pages} {057002} (\bibinfo {year}
  {2009})}\BibitemShut {NoStop}%
\bibitem [{\citenamefont {Tan}\ \emph {et~al.}(2013)\citenamefont {Tan},
  \citenamefont {Zhang}, \citenamefont {Xia}, \citenamefont {Ye}, ,
  \citenamefont {Chen}, \citenamefont {Xie}, \citenamefont {Peng},
  \citenamefont {Xu}, \citenamefont {Qin~Fan}, \citenamefont {Jiang},
  \citenamefont {Zhang}, \citenamefont {Lai}, \citenamefont {Xiang},
  \citenamefont {Hu}, \citenamefont {Xie},\ and\ \citenamefont
  {Feng}}]{Tan13NatMat}%
  \BibitemOpen
  \bibfield  {author} {\bibinfo {author} {\bibfnamefont {S.}~\bibnamefont
  {Tan}}, \bibinfo {author} {\bibfnamefont {Y.}~\bibnamefont {Zhang}}, \bibinfo
  {author} {\bibfnamefont {M.}~\bibnamefont {Xia}}, \bibinfo {author}
  {\bibfnamefont {Z.}~\bibnamefont {Ye}}, , \bibinfo {author} {\bibfnamefont
  {F.}~\bibnamefont {Chen}}, \bibinfo {author} {\bibfnamefont {X.}~\bibnamefont
  {Xie}}, \bibinfo {author} {\bibfnamefont {R.}~\bibnamefont {Peng}}, \bibinfo
  {author} {\bibfnamefont {D.}~\bibnamefont {Xu}}, \bibinfo {author}
  {\bibfnamefont {H.~X.}\ \bibnamefont {Qin~Fan}}, \bibinfo {author}
  {\bibfnamefont {J.}~\bibnamefont {Jiang}}, \bibinfo {author} {\bibfnamefont
  {T.}~\bibnamefont {Zhang}}, \bibinfo {author} {\bibfnamefont
  {X.}~\bibnamefont {Lai}}, \bibinfo {author} {\bibfnamefont {T.}~\bibnamefont
  {Xiang}}, \bibinfo {author} {\bibfnamefont {J.}~\bibnamefont {Hu}}, \bibinfo
  {author} {\bibfnamefont {B.}~\bibnamefont {Xie}}, \ and\ \bibinfo {author}
  {\bibfnamefont {D.}~\bibnamefont {Feng}},\ }\href@noop {} {\bibfield
  {journal} {\bibinfo  {journal} {Nat. Mater.}\ }\textbf {\bibinfo {volume}
  {12}},\ \bibinfo {pages} {634} (\bibinfo {year} {2013})}\BibitemShut
  {NoStop}%
\bibitem [{\citenamefont {Shimojima}\ \emph {et~al.}(2014)\citenamefont
  {Shimojima}, \citenamefont {Suzuki}, \citenamefont {Sonobe}, \citenamefont
  {Nakamura}, \citenamefont {Sakano}, \citenamefont {Omachi}, \citenamefont
  {Yoshioka}, \citenamefont {Kuwata-Gonokami}, \citenamefont {Ono},
  \citenamefont {Kumigashira}, \citenamefont {B\"ohmer}, \citenamefont {Hardy},
  \citenamefont {Wolf}, \citenamefont {Meingast}, \citenamefont {L\"ohneysen},
  \citenamefont {Ikeda},\ and\ \citenamefont {Ishizaka}}]{Shimojima14PRB}%
  \BibitemOpen
  \bibfield  {author} {\bibinfo {author} {\bibfnamefont {T.}~\bibnamefont
  {Shimojima}}, \bibinfo {author} {\bibfnamefont {Y.}~\bibnamefont {Suzuki}},
  \bibinfo {author} {\bibfnamefont {T.}~\bibnamefont {Sonobe}}, \bibinfo
  {author} {\bibfnamefont {A.}~\bibnamefont {Nakamura}}, \bibinfo {author}
  {\bibfnamefont {M.}~\bibnamefont {Sakano}}, \bibinfo {author} {\bibfnamefont
  {J.}~\bibnamefont {Omachi}}, \bibinfo {author} {\bibfnamefont
  {K.}~\bibnamefont {Yoshioka}}, \bibinfo {author} {\bibfnamefont
  {M.}~\bibnamefont {Kuwata-Gonokami}}, \bibinfo {author} {\bibfnamefont
  {K.}~\bibnamefont {Ono}}, \bibinfo {author} {\bibfnamefont {H.}~\bibnamefont
  {Kumigashira}}, \bibinfo {author} {\bibfnamefont {A.~E.}\ \bibnamefont
  {B\"ohmer}}, \bibinfo {author} {\bibfnamefont {F.}~\bibnamefont {Hardy}},
  \bibinfo {author} {\bibfnamefont {T.}~\bibnamefont {Wolf}}, \bibinfo {author}
  {\bibfnamefont {C.}~\bibnamefont {Meingast}}, \bibinfo {author}
  {\bibfnamefont {H.~v.}\ \bibnamefont {L\"ohneysen}}, \bibinfo {author}
  {\bibfnamefont {H.}~\bibnamefont {Ikeda}}, \ and\ \bibinfo {author}
  {\bibfnamefont {K.}~\bibnamefont {Ishizaka}},\ }\href {\doibase
  10.1103/PhysRevB.90.121111} {\bibfield  {journal} {\bibinfo  {journal} {Phys.
  Rev. B}\ }\textbf {\bibinfo {volume} {90}},\ \bibinfo {pages} {121111}
  (\bibinfo {year} {2014})}\BibitemShut {NoStop}%
\bibitem [{\citenamefont {Nakayama}\ \emph {et~al.}(2014)\citenamefont
  {Nakayama}, \citenamefont {Miyata}, \citenamefont {Phan}, \citenamefont
  {Sato}, \citenamefont {Tanabe}, \citenamefont {Urata}, \citenamefont
  {Tanigaki},\ and\ \citenamefont {Takahashi}}]{Nakayama14PRL}%
  \BibitemOpen
  \bibfield  {author} {\bibinfo {author} {\bibfnamefont {K.}~\bibnamefont
  {Nakayama}}, \bibinfo {author} {\bibfnamefont {Y.}~\bibnamefont {Miyata}},
  \bibinfo {author} {\bibfnamefont {G.~N.}\ \bibnamefont {Phan}}, \bibinfo
  {author} {\bibfnamefont {T.}~\bibnamefont {Sato}}, \bibinfo {author}
  {\bibfnamefont {Y.}~\bibnamefont {Tanabe}}, \bibinfo {author} {\bibfnamefont
  {T.}~\bibnamefont {Urata}}, \bibinfo {author} {\bibfnamefont
  {K.}~\bibnamefont {Tanigaki}}, \ and\ \bibinfo {author} {\bibfnamefont
  {T.}~\bibnamefont {Takahashi}},\ }\href {\doibase
  10.1103/PhysRevLett.113.237001} {\bibfield  {journal} {\bibinfo  {journal}
  {Phys. Rev. Lett.}\ }\textbf {\bibinfo {volume} {113}},\ \bibinfo {pages}
  {237001} (\bibinfo {year} {2014})}\BibitemShut {NoStop}%
\bibitem [{\citenamefont {Imai}\ \emph {et~al.}(2009)\citenamefont {Imai},
  \citenamefont {Ahilan}, \citenamefont {Ning}, \citenamefont {McQueen},\ and\
  \citenamefont {Cava}}]{Imai09PRL}%
  \BibitemOpen
  \bibfield  {author} {\bibinfo {author} {\bibfnamefont {T.}~\bibnamefont
  {Imai}}, \bibinfo {author} {\bibfnamefont {K.}~\bibnamefont {Ahilan}},
  \bibinfo {author} {\bibfnamefont {F.~L.}\ \bibnamefont {Ning}}, \bibinfo
  {author} {\bibfnamefont {T.~M.}\ \bibnamefont {McQueen}}, \ and\ \bibinfo
  {author} {\bibfnamefont {R.~J.}\ \bibnamefont {Cava}},\ }\href {\doibase
  10.1103/PhysRevLett.102.177005} {\bibfield  {journal} {\bibinfo  {journal}
  {Phys. Rev. Lett.}\ }\textbf {\bibinfo {volume} {102}},\ \bibinfo {pages}
  {177005} (\bibinfo {year} {2009})}\BibitemShut {NoStop}%
\bibitem [{\citenamefont {Baek}\ \emph {et~al.}(2014)\citenamefont {Baek},
  \citenamefont {Efremov}, \citenamefont {Ok}, \citenamefont {Kim},
  \citenamefont {van~den Brink},\ and\ \citenamefont {B\"uchner}}]{Baek14nmat}%
  \BibitemOpen
  \bibfield  {author} {\bibinfo {author} {\bibfnamefont {S.-H.}\ \bibnamefont
  {Baek}}, \bibinfo {author} {\bibfnamefont {D.~V.}\ \bibnamefont {Efremov}},
  \bibinfo {author} {\bibfnamefont {J.~M.}\ \bibnamefont {Ok}}, \bibinfo
  {author} {\bibfnamefont {J.~S.}\ \bibnamefont {Kim}}, \bibinfo {author}
  {\bibfnamefont {J.}~\bibnamefont {van~den Brink}}, \ and\ \bibinfo {author}
  {\bibfnamefont {B.}~\bibnamefont {B\"uchner}},\ }\href@noop {} {\bibfield
  {journal} {\bibinfo  {journal} {Nat. Mater.}\ }\textbf {\bibinfo {volume}
  {14}},\ \bibinfo {pages} {210} (\bibinfo {year} {2014})}\BibitemShut
  {NoStop}%
\bibitem [{\citenamefont {B\"ohmer}\ \emph {et~al.}(2015)\citenamefont
  {B\"ohmer}, \citenamefont {Arai}, \citenamefont {Hardy}, \citenamefont
  {Hattori}, \citenamefont {Iye}, \citenamefont {Wolf}, \citenamefont
  {L\"ohneysen}, \citenamefont {Ishida},\ and\ \citenamefont
  {Meingast}}]{Bohmer15PRL}%
  \BibitemOpen
  \bibfield  {author} {\bibinfo {author} {\bibfnamefont {A.~E.}\ \bibnamefont
  {B\"ohmer}}, \bibinfo {author} {\bibfnamefont {T.}~\bibnamefont {Arai}},
  \bibinfo {author} {\bibfnamefont {F.}~\bibnamefont {Hardy}}, \bibinfo
  {author} {\bibfnamefont {T.}~\bibnamefont {Hattori}}, \bibinfo {author}
  {\bibfnamefont {T.}~\bibnamefont {Iye}}, \bibinfo {author} {\bibfnamefont
  {T.}~\bibnamefont {Wolf}}, \bibinfo {author} {\bibfnamefont {H.~v.}\
  \bibnamefont {L\"ohneysen}}, \bibinfo {author} {\bibfnamefont
  {K.}~\bibnamefont {Ishida}}, \ and\ \bibinfo {author} {\bibfnamefont
  {C.}~\bibnamefont {Meingast}},\ }\href {\doibase
  10.1103/PhysRevLett.114.027001} {\bibfield  {journal} {\bibinfo  {journal}
  {Phys. Rev. Lett.}\ }\textbf {\bibinfo {volume} {114}},\ \bibinfo {pages}
  {027001} (\bibinfo {year} {2015})}\BibitemShut {NoStop}%
\bibitem [{\citenamefont {Rahn}\ \emph {et~al.}(2015)\citenamefont {Rahn},
  \citenamefont {Ewings}, \citenamefont {Sedlmaier}, \citenamefont {Clarke},\
  and\ \citenamefont {Boothroyd}}]{Rahn15PRB}%
  \BibitemOpen
  \bibfield  {author} {\bibinfo {author} {\bibfnamefont {M.~C.}\ \bibnamefont
  {Rahn}}, \bibinfo {author} {\bibfnamefont {R.~A.}\ \bibnamefont {Ewings}},
  \bibinfo {author} {\bibfnamefont {S.~J.}\ \bibnamefont {Sedlmaier}}, \bibinfo
  {author} {\bibfnamefont {S.~J.}\ \bibnamefont {Clarke}}, \ and\ \bibinfo
  {author} {\bibfnamefont {A.~T.}\ \bibnamefont {Boothroyd}},\ }\href {\doibase
  10.1103/PhysRevB.91.180501} {\bibfield  {journal} {\bibinfo  {journal} {Phys.
  Rev. B}\ }\textbf {\bibinfo {volume} {91}},\ \bibinfo {pages} {180501}
  (\bibinfo {year} {2015})}\BibitemShut {NoStop}%
\bibitem [{\citenamefont {Wang}\ \emph {et~al.}(2016)\citenamefont {Wang},
  \citenamefont {Shen}, \citenamefont {Pan}, \citenamefont {Hao}, \citenamefont
  {Ma}, \citenamefont {Zhou}, \citenamefont {Steffens}, \citenamefont
  {Schmalzl}, \citenamefont {Forrest}, \citenamefont {Abdel-Hafiez},
  \citenamefont {Chen}, \citenamefont {Chareev}, \citenamefont {Vasiliev},
  \citenamefont {Bourges}, \citenamefont {Sidis}, \citenamefont {Cao},\ and\
  \citenamefont {Zhao}}]{Wang15NatMater}%
  \BibitemOpen
  \bibfield  {author} {\bibinfo {author} {\bibfnamefont {Q.}~\bibnamefont
  {Wang}}, \bibinfo {author} {\bibfnamefont {Y.}~\bibnamefont {Shen}}, \bibinfo
  {author} {\bibfnamefont {B.}~\bibnamefont {Pan}}, \bibinfo {author}
  {\bibfnamefont {Y.}~\bibnamefont {Hao}}, \bibinfo {author} {\bibfnamefont
  {M.}~\bibnamefont {Ma}}, \bibinfo {author} {\bibfnamefont {F.}~\bibnamefont
  {Zhou}}, \bibinfo {author} {\bibfnamefont {P.}~\bibnamefont {Steffens}},
  \bibinfo {author} {\bibfnamefont {K.}~\bibnamefont {Schmalzl}}, \bibinfo
  {author} {\bibfnamefont {T.~R.}\ \bibnamefont {Forrest}}, \bibinfo {author}
  {\bibfnamefont {M.}~\bibnamefont {Abdel-Hafiez}}, \bibinfo {author}
  {\bibfnamefont {X.}~\bibnamefont {Chen}}, \bibinfo {author} {\bibfnamefont
  {D.~A.}\ \bibnamefont {Chareev}}, \bibinfo {author} {\bibfnamefont {A.~N.}\
  \bibnamefont {Vasiliev}}, \bibinfo {author} {\bibfnamefont {P.}~\bibnamefont
  {Bourges}}, \bibinfo {author} {\bibfnamefont {Y.}~\bibnamefont {Sidis}},
  \bibinfo {author} {\bibfnamefont {H.}~\bibnamefont {Cao}}, \ and\ \bibinfo
  {author} {\bibfnamefont {J.}~\bibnamefont {Zhao}},\ }\href
  {http://dx.doi.org/10.1038/nmat4492} {\bibfield  {journal} {\bibinfo
  {journal} {Nat Mater}\ }\textbf {\bibinfo {volume} {15}},\ \bibinfo {pages}
  {159} (\bibinfo {year} {2016})}\BibitemShut {NoStop}%
\bibitem [{\citenamefont {Bendele}\ \emph {et~al.}(2010)\citenamefont
  {Bendele}, \citenamefont {Amato}, \citenamefont {Conder}, \citenamefont
  {Elender}, \citenamefont {Keller}, \citenamefont {Klauss}, \citenamefont
  {Luetkens}, \citenamefont {Pomjakushina}, \citenamefont {Raselli},\ and\
  \citenamefont {Khasanov}}]{Bendele10PRL}%
  \BibitemOpen
  \bibfield  {author} {\bibinfo {author} {\bibfnamefont {M.}~\bibnamefont
  {Bendele}}, \bibinfo {author} {\bibfnamefont {A.}~\bibnamefont {Amato}},
  \bibinfo {author} {\bibfnamefont {K.}~\bibnamefont {Conder}}, \bibinfo
  {author} {\bibfnamefont {M.}~\bibnamefont {Elender}}, \bibinfo {author}
  {\bibfnamefont {H.}~\bibnamefont {Keller}}, \bibinfo {author} {\bibfnamefont
  {H.-H.}\ \bibnamefont {Klauss}}, \bibinfo {author} {\bibfnamefont
  {H.}~\bibnamefont {Luetkens}}, \bibinfo {author} {\bibfnamefont
  {E.}~\bibnamefont {Pomjakushina}}, \bibinfo {author} {\bibfnamefont
  {A.}~\bibnamefont {Raselli}}, \ and\ \bibinfo {author} {\bibfnamefont
  {R.}~\bibnamefont {Khasanov}},\ }\href {\doibase
  10.1103/PhysRevLett.104.087003} {\bibfield  {journal} {\bibinfo  {journal}
  {Phys. Rev. Lett.}\ }\textbf {\bibinfo {volume} {104}},\ \bibinfo {pages}
  {087003} (\bibinfo {year} {2010})}\BibitemShut {NoStop}%
\bibitem [{\citenamefont {Bendele}\ \emph {et~al.}(2012)\citenamefont
  {Bendele}, \citenamefont {Ichsanow}, \citenamefont {Pashkevich},
  \citenamefont {Keller}, \citenamefont {Str\"assle}, \citenamefont {Gusev},
  \citenamefont {Pomjakushina}, \citenamefont {Conder}, \citenamefont
  {Khasanov},\ and\ \citenamefont {Keller}}]{Bendele12PRB}%
  \BibitemOpen
  \bibfield  {author} {\bibinfo {author} {\bibfnamefont {M.}~\bibnamefont
  {Bendele}}, \bibinfo {author} {\bibfnamefont {A.}~\bibnamefont {Ichsanow}},
  \bibinfo {author} {\bibfnamefont {Y.}~\bibnamefont {Pashkevich}}, \bibinfo
  {author} {\bibfnamefont {L.}~\bibnamefont {Keller}}, \bibinfo {author}
  {\bibfnamefont {T.}~\bibnamefont {Str\"assle}}, \bibinfo {author}
  {\bibfnamefont {A.}~\bibnamefont {Gusev}}, \bibinfo {author} {\bibfnamefont
  {E.}~\bibnamefont {Pomjakushina}}, \bibinfo {author} {\bibfnamefont
  {K.}~\bibnamefont {Conder}}, \bibinfo {author} {\bibfnamefont
  {R.}~\bibnamefont {Khasanov}}, \ and\ \bibinfo {author} {\bibfnamefont
  {H.}~\bibnamefont {Keller}},\ }\href {\doibase 10.1103/PhysRevB.85.064517}
  {\bibfield  {journal} {\bibinfo  {journal} {Phys. Rev. B}\ }\textbf {\bibinfo
  {volume} {85}},\ \bibinfo {pages} {064517} (\bibinfo {year}
  {2012})}\BibitemShut {NoStop}%
\bibitem [{\citenamefont {Terashima}\ \emph {et~al.}(2015)\citenamefont
  {Terashima}, \citenamefont {Kikugawa}, \citenamefont {Kasahara},
  \citenamefont {Watashige}, \citenamefont {Shibauchi}, \citenamefont
  {Matsuda}, \citenamefont {Wolf}, \citenamefont {B\"ohmer}, \citenamefont
  {Hardy}, \citenamefont {Meingast}, \citenamefont {v.~L\"ohneysen},\ and\
  \citenamefont {Uji}}]{Terashima15JPSJ}%
  \BibitemOpen
  \bibfield  {author} {\bibinfo {author} {\bibfnamefont {T.}~\bibnamefont
  {Terashima}}, \bibinfo {author} {\bibfnamefont {N.}~\bibnamefont {Kikugawa}},
  \bibinfo {author} {\bibfnamefont {S.}~\bibnamefont {Kasahara}}, \bibinfo
  {author} {\bibfnamefont {T.}~\bibnamefont {Watashige}}, \bibinfo {author}
  {\bibfnamefont {T.}~\bibnamefont {Shibauchi}}, \bibinfo {author}
  {\bibfnamefont {Y.}~\bibnamefont {Matsuda}}, \bibinfo {author} {\bibfnamefont
  {T.}~\bibnamefont {Wolf}}, \bibinfo {author} {\bibfnamefont {A.~E.}\
  \bibnamefont {B\"ohmer}}, \bibinfo {author} {\bibfnamefont {F.}~\bibnamefont
  {Hardy}}, \bibinfo {author} {\bibfnamefont {C.}~\bibnamefont {Meingast}},
  \bibinfo {author} {\bibfnamefont {H.}~\bibnamefont {v.~L\"ohneysen}}, \ and\
  \bibinfo {author} {\bibfnamefont {S.}~\bibnamefont {Uji}},\ }\href {\doibase
  10.7566/JPSJ.84.063701} {\bibfield  {journal} {\bibinfo  {journal} {J. Phys.
  Soc. Jpn.}\ }\textbf {\bibinfo {volume} {84}},\ \bibinfo {pages} {063701}
  (\bibinfo {year} {2015})}\BibitemShut {NoStop}%
\bibitem [{\citenamefont {Yamase}\ and\ \citenamefont
  {Zeyher}(2015)}]{Yamase15NJP}%
  \BibitemOpen
  \bibfield  {author} {\bibinfo {author} {\bibfnamefont {H.}~\bibnamefont
  {Yamase}}\ and\ \bibinfo {author} {\bibfnamefont {R.}~\bibnamefont
  {Zeyher}},\ }\href {http://stacks.iop.org/1367-2630/17/i=7/a=073030}
  {\bibfield  {journal} {\bibinfo  {journal} {New J. Phys.}\ }\textbf {\bibinfo
  {volume} {17}},\ \bibinfo {pages} {073030} (\bibinfo {year}
  {2015})}\BibitemShut {NoStop}%
\bibitem [{\citenamefont {Essenberger}\ \emph {et~al.}(2012)\citenamefont
  {Essenberger}, \citenamefont {Buczek}, \citenamefont {Ernst}, \citenamefont
  {Sandratskii},\ and\ \citenamefont {Gross}}]{Essenberger12PRB}%
  \BibitemOpen
  \bibfield  {author} {\bibinfo {author} {\bibfnamefont {F.}~\bibnamefont
  {Essenberger}}, \bibinfo {author} {\bibfnamefont {P.}~\bibnamefont {Buczek}},
  \bibinfo {author} {\bibfnamefont {A.}~\bibnamefont {Ernst}}, \bibinfo
  {author} {\bibfnamefont {L.}~\bibnamefont {Sandratskii}}, \ and\ \bibinfo
  {author} {\bibfnamefont {E.~K.~U.}\ \bibnamefont {Gross}},\ }\href {\doibase
  10.1103/PhysRevB.86.060412} {\bibfield  {journal} {\bibinfo  {journal} {Phys.
  Rev. B}\ }\textbf {\bibinfo {volume} {86}},\ \bibinfo {pages} {060412}
  (\bibinfo {year} {2012})}\BibitemShut {NoStop}%
\bibitem [{\citenamefont {Lischner}\ \emph {et~al.}(2015)\citenamefont
  {Lischner}, \citenamefont {Bazhirov}, \citenamefont {MacDonald},
  \citenamefont {Cohen},\ and\ \citenamefont {Louie}}]{Lischner15PRB}%
  \BibitemOpen
  \bibfield  {author} {\bibinfo {author} {\bibfnamefont {J.}~\bibnamefont
  {Lischner}}, \bibinfo {author} {\bibfnamefont {T.}~\bibnamefont {Bazhirov}},
  \bibinfo {author} {\bibfnamefont {A.~H.}\ \bibnamefont {MacDonald}}, \bibinfo
  {author} {\bibfnamefont {M.~L.}\ \bibnamefont {Cohen}}, \ and\ \bibinfo
  {author} {\bibfnamefont {S.~G.}\ \bibnamefont {Louie}},\ }\href {\doibase
  10.1103/PhysRevB.91.020502} {\bibfield  {journal} {\bibinfo  {journal} {Phys.
  Rev. B}\ }\textbf {\bibinfo {volume} {91}},\ \bibinfo {pages} {020502}
  (\bibinfo {year} {2015})}\BibitemShut {NoStop}%
\bibitem [{\citenamefont {Chubukov}\ \emph {et~al.}(2015)\citenamefont
  {Chubukov}, \citenamefont {Fernandes},\ and\ \citenamefont
  {Schmalian}}]{Chubukov15PRB}%
  \BibitemOpen
  \bibfield  {author} {\bibinfo {author} {\bibfnamefont {A.~V.}\ \bibnamefont
  {Chubukov}}, \bibinfo {author} {\bibfnamefont {R.~M.}\ \bibnamefont
  {Fernandes}}, \ and\ \bibinfo {author} {\bibfnamefont {J.}~\bibnamefont
  {Schmalian}},\ }\href {\doibase 10.1103/PhysRevB.91.201105} {\bibfield
  {journal} {\bibinfo  {journal} {Phys. Rev. B}\ }\textbf {\bibinfo {volume}
  {91}},\ \bibinfo {pages} {201105} (\bibinfo {year} {2015})}\BibitemShut
  {NoStop}%
\bibitem [{\citenamefont {Mukherjee}\ \emph {et~al.}(2015)\citenamefont
  {Mukherjee}, \citenamefont {Kreisel}, \citenamefont {Hirschfeld},\ and\
  \citenamefont {Andersen}}]{Mukherjee15PRL}%
  \BibitemOpen
  \bibfield  {author} {\bibinfo {author} {\bibfnamefont {S.}~\bibnamefont
  {Mukherjee}}, \bibinfo {author} {\bibfnamefont {A.}~\bibnamefont {Kreisel}},
  \bibinfo {author} {\bibfnamefont {P.~J.}\ \bibnamefont {Hirschfeld}}, \ and\
  \bibinfo {author} {\bibfnamefont {B.~M.}\ \bibnamefont {Andersen}},\ }\href
  {\doibase 10.1103/PhysRevLett.115.026402} {\bibfield  {journal} {\bibinfo
  {journal} {Phys. Rev. Lett.}\ }\textbf {\bibinfo {volume} {115}},\ \bibinfo
  {pages} {026402} (\bibinfo {year} {2015})}\BibitemShut {NoStop}%
\bibitem [{\citenamefont {Glasbrenner}\ \emph {et~al.}(2015)\citenamefont
  {Glasbrenner}, \citenamefont {Mazin}, \citenamefont {Jeschke}, \citenamefont
  {Hirschfeld}, \citenamefont {Fernandes},\ and\ \citenamefont
  {Valenti}}]{Glasbrenner15NatPhys}%
  \BibitemOpen
  \bibfield  {author} {\bibinfo {author} {\bibfnamefont {J.~K.}\ \bibnamefont
  {Glasbrenner}}, \bibinfo {author} {\bibfnamefont {I.~I.}\ \bibnamefont
  {Mazin}}, \bibinfo {author} {\bibfnamefont {H.~O.}\ \bibnamefont {Jeschke}},
  \bibinfo {author} {\bibfnamefont {P.~J.}\ \bibnamefont {Hirschfeld}},
  \bibinfo {author} {\bibfnamefont {R.~M.}\ \bibnamefont {Fernandes}}, \ and\
  \bibinfo {author} {\bibfnamefont {R.}~\bibnamefont {Valenti}},\ }\href
  {http://dx.doi.org/10.1038/nphys3434} {\bibfield  {journal} {\bibinfo
  {journal} {Nat Phys}\ }\textbf {\bibinfo {volume} {11}},\ \bibinfo {pages}
  {953} (\bibinfo {year} {2015})}\BibitemShut {NoStop}%
\bibitem [{\citenamefont {Hirayama}\ \emph {et~al.}(2015)\citenamefont
  {Hirayama}, \citenamefont {Misawa}, \citenamefont {Miyake},\ and\
  \citenamefont {Imada}}]{Hirayama15JPSJ}%
  \BibitemOpen
  \bibfield  {author} {\bibinfo {author} {\bibfnamefont {M.}~\bibnamefont
  {Hirayama}}, \bibinfo {author} {\bibfnamefont {T.}~\bibnamefont {Misawa}},
  \bibinfo {author} {\bibfnamefont {T.}~\bibnamefont {Miyake}}, \ and\ \bibinfo
  {author} {\bibfnamefont {M.}~\bibnamefont {Imada}},\ }\href {\doibase
  10.7566/JPSJ.84.093703} {\bibfield  {journal} {\bibinfo  {journal} {J. Phys.
  Soc. Jpn.}\ }\textbf {\bibinfo {volume} {84}},\ \bibinfo {pages} {093703}
  (\bibinfo {year} {2015})}\BibitemShut {NoStop}%
\bibitem [{\citenamefont {Leonov}\ \emph {et~al.}(2015)\citenamefont {Leonov},
  \citenamefont {Skornyakov}, \citenamefont {Anisimov},\ and\ \citenamefont
  {Vollhardt}}]{Leonov15PRL}%
  \BibitemOpen
  \bibfield  {author} {\bibinfo {author} {\bibfnamefont {I.}~\bibnamefont
  {Leonov}}, \bibinfo {author} {\bibfnamefont {S.~L.}\ \bibnamefont
  {Skornyakov}}, \bibinfo {author} {\bibfnamefont {V.~I.}\ \bibnamefont
  {Anisimov}}, \ and\ \bibinfo {author} {\bibfnamefont {D.}~\bibnamefont
  {Vollhardt}},\ }\href {\doibase 10.1103/PhysRevLett.115.106402} {\bibfield
  {journal} {\bibinfo  {journal} {Phys. Rev. Lett.}\ }\textbf {\bibinfo
  {volume} {115}},\ \bibinfo {pages} {106402} (\bibinfo {year}
  {2015})}\BibitemShut {NoStop}%
\bibitem [{\citenamefont {Wang}\ \emph {et~al.}(2015)\citenamefont {Wang},
  \citenamefont {Kivelson},\ and\ \citenamefont {Lee}}]{Wang15NatPhys}%
  \BibitemOpen
  \bibfield  {author} {\bibinfo {author} {\bibfnamefont {F.}~\bibnamefont
  {Wang}}, \bibinfo {author} {\bibfnamefont {S.~A.}\ \bibnamefont {Kivelson}},
  \ and\ \bibinfo {author} {\bibfnamefont {D.-H.}\ \bibnamefont {Lee}},\ }\href
  {http://dx.doi.org/10.1038/nphys3456} {\bibfield  {journal} {\bibinfo
  {journal} {Nat Phys}\ }\textbf {\bibinfo {volume} {11}},\ \bibinfo {pages}
  {959} (\bibinfo {year} {2015})}\BibitemShut {NoStop}%
\bibitem [{\citenamefont {Yu}\ and\ \citenamefont {Si}(2015)}]{Rong15PRL}%
  \BibitemOpen
  \bibfield  {author} {\bibinfo {author} {\bibfnamefont {R.}~\bibnamefont
  {Yu}}\ and\ \bibinfo {author} {\bibfnamefont {Q.}~\bibnamefont {Si}},\ }\href
  {\doibase 10.1103/PhysRevLett.115.116401} {\bibfield  {journal} {\bibinfo
  {journal} {Phys. Rev. Lett.}\ }\textbf {\bibinfo {volume} {115}},\ \bibinfo
  {pages} {116401} (\bibinfo {year} {2015})}\BibitemShut {NoStop}%
\bibitem [{\citenamefont {Jiang}\ \emph {et~al.}(2015)\citenamefont {Jiang},
  \citenamefont {Hu}, \citenamefont {Ding},\ and\ \citenamefont
  {Wang}}]{Jiang15condmat}%
  \BibitemOpen
  \bibfield  {author} {\bibinfo {author} {\bibfnamefont {K.}~\bibnamefont
  {Jiang}}, \bibinfo {author} {\bibfnamefont {J.}~\bibnamefont {Hu}}, \bibinfo
  {author} {\bibfnamefont {H.}~\bibnamefont {Ding}}, \ and\ \bibinfo {author}
  {\bibfnamefont {Z.}~\bibnamefont {Wang}},\ }\href@noop {} {\bibfield
  {journal} {\bibinfo  {journal} {arXiv:1508.00588}\ } (\bibinfo {year}
  {2015})}\BibitemShut {NoStop}%
\bibitem [{\citenamefont {Yamakawa}\ \emph {et~al.}(2015)\citenamefont
  {Yamakawa}, \citenamefont {Onari},\ and\ \citenamefont
  {Kontani}}]{Yamakawa15condmat}%
  \BibitemOpen
  \bibfield  {author} {\bibinfo {author} {\bibfnamefont {Y.}~\bibnamefont
  {Yamakawa}}, \bibinfo {author} {\bibfnamefont {S.}~\bibnamefont {Onari}}, \
  and\ \bibinfo {author} {\bibfnamefont {H.}~\bibnamefont {Kontani}},\
  }\href@noop {} {\bibfield  {journal} {\bibinfo  {journal} {arXiv:1509.01161}\
  } (\bibinfo {year} {2015})}\BibitemShut {NoStop}%
\bibitem [{\citenamefont {Terashima}\ \emph
  {et~al.}(2014{\natexlab{a}})\citenamefont {Terashima}, \citenamefont
  {Kikugawa}, \citenamefont {Kiswandhi}, \citenamefont {Choi}, \citenamefont
  {Brooks}, \citenamefont {Kasahara}, \citenamefont {Watashige}, \citenamefont
  {Ikeda}, \citenamefont {Shibauchi}, \citenamefont {Matsuda}, \citenamefont
  {Wolf}, \citenamefont {B\"ohmer}, \citenamefont {Hardy}, \citenamefont
  {Meingast}, \citenamefont {L\"ohneysen}, \citenamefont {Suzuki},
  \citenamefont {Arita},\ and\ \citenamefont {Uji}}]{Terashima14PRB}%
  \BibitemOpen
  \bibfield  {author} {\bibinfo {author} {\bibfnamefont {T.}~\bibnamefont
  {Terashima}}, \bibinfo {author} {\bibfnamefont {N.}~\bibnamefont {Kikugawa}},
  \bibinfo {author} {\bibfnamefont {A.}~\bibnamefont {Kiswandhi}}, \bibinfo
  {author} {\bibfnamefont {E.-S.}\ \bibnamefont {Choi}}, \bibinfo {author}
  {\bibfnamefont {J.~S.}\ \bibnamefont {Brooks}}, \bibinfo {author}
  {\bibfnamefont {S.}~\bibnamefont {Kasahara}}, \bibinfo {author}
  {\bibfnamefont {T.}~\bibnamefont {Watashige}}, \bibinfo {author}
  {\bibfnamefont {H.}~\bibnamefont {Ikeda}}, \bibinfo {author} {\bibfnamefont
  {T.}~\bibnamefont {Shibauchi}}, \bibinfo {author} {\bibfnamefont
  {Y.}~\bibnamefont {Matsuda}}, \bibinfo {author} {\bibfnamefont
  {T.}~\bibnamefont {Wolf}}, \bibinfo {author} {\bibfnamefont {A.~E.}\
  \bibnamefont {B\"ohmer}}, \bibinfo {author} {\bibfnamefont {F.}~\bibnamefont
  {Hardy}}, \bibinfo {author} {\bibfnamefont {C.}~\bibnamefont {Meingast}},
  \bibinfo {author} {\bibfnamefont {H.~v.}\ \bibnamefont {L\"ohneysen}},
  \bibinfo {author} {\bibfnamefont {M.-T.}\ \bibnamefont {Suzuki}}, \bibinfo
  {author} {\bibfnamefont {R.}~\bibnamefont {Arita}}, \ and\ \bibinfo {author}
  {\bibfnamefont {S.}~\bibnamefont {Uji}},\ }\href {\doibase
  10.1103/PhysRevB.90.144517} {\bibfield  {journal} {\bibinfo  {journal} {Phys.
  Rev. B}\ }\textbf {\bibinfo {volume} {90}},\ \bibinfo {pages} {144517}
  (\bibinfo {year} {2014}{\natexlab{a}})}\BibitemShut {NoStop}%
\bibitem [{\citenamefont {Audouard}\ \emph {et~al.}(2015)\citenamefont
  {Audouard}, \citenamefont {Duc}, \citenamefont {Drigo}, \citenamefont
  {Toulemonde}, \citenamefont {Karlsson}, \citenamefont {Strobel},\ and\
  \citenamefont {Sulpice}}]{Audouard15EPL}%
  \BibitemOpen
  \bibfield  {author} {\bibinfo {author} {\bibfnamefont {A.}~\bibnamefont
  {Audouard}}, \bibinfo {author} {\bibfnamefont {F.}~\bibnamefont {Duc}},
  \bibinfo {author} {\bibfnamefont {L.}~\bibnamefont {Drigo}}, \bibinfo
  {author} {\bibfnamefont {P.}~\bibnamefont {Toulemonde}}, \bibinfo {author}
  {\bibfnamefont {S.}~\bibnamefont {Karlsson}}, \bibinfo {author}
  {\bibfnamefont {P.}~\bibnamefont {Strobel}}, \ and\ \bibinfo {author}
  {\bibfnamefont {A.}~\bibnamefont {Sulpice}},\ }\href
  {http://stacks.iop.org/0295-5075/109/i=2/a=27003} {\bibfield  {journal}
  {\bibinfo  {journal} {EPL (Europhysics Letters)}\ }\textbf {\bibinfo {volume}
  {109}},\ \bibinfo {pages} {27003} (\bibinfo {year} {2015})}\BibitemShut
  {NoStop}%
\bibitem [{\citenamefont {Watson}\ \emph
  {et~al.}(2015{\natexlab{a}})\citenamefont {Watson}, \citenamefont {Kim},
  \citenamefont {Haghighirad}, \citenamefont {Davies}, \citenamefont
  {McCollam}, \citenamefont {Narayanan}, \citenamefont {Blake}, \citenamefont
  {Chen}, \citenamefont {Ghannadzadeh}, \citenamefont {Schofield},
  \citenamefont {Hoesch}, \citenamefont {Meingast}, \citenamefont {Wolf},\ and\
  \citenamefont {Coldea}}]{WatsonPRB15}%
  \BibitemOpen
  \bibfield  {author} {\bibinfo {author} {\bibfnamefont {M.~D.}\ \bibnamefont
  {Watson}}, \bibinfo {author} {\bibfnamefont {T.~K.}\ \bibnamefont {Kim}},
  \bibinfo {author} {\bibfnamefont {A.~A.}\ \bibnamefont {Haghighirad}},
  \bibinfo {author} {\bibfnamefont {N.~R.}\ \bibnamefont {Davies}}, \bibinfo
  {author} {\bibfnamefont {A.}~\bibnamefont {McCollam}}, \bibinfo {author}
  {\bibfnamefont {A.}~\bibnamefont {Narayanan}}, \bibinfo {author}
  {\bibfnamefont {S.~F.}\ \bibnamefont {Blake}}, \bibinfo {author}
  {\bibfnamefont {Y.~L.}\ \bibnamefont {Chen}}, \bibinfo {author}
  {\bibfnamefont {S.}~\bibnamefont {Ghannadzadeh}}, \bibinfo {author}
  {\bibfnamefont {A.~J.}\ \bibnamefont {Schofield}}, \bibinfo {author}
  {\bibfnamefont {M.}~\bibnamefont {Hoesch}}, \bibinfo {author} {\bibfnamefont
  {C.}~\bibnamefont {Meingast}}, \bibinfo {author} {\bibfnamefont
  {T.}~\bibnamefont {Wolf}}, \ and\ \bibinfo {author} {\bibfnamefont {A.~I.}\
  \bibnamefont {Coldea}},\ }\href {\doibase 10.1103/PhysRevB.91.155106}
  {\bibfield  {journal} {\bibinfo  {journal} {Phys. Rev. B}\ }\textbf {\bibinfo
  {volume} {91}},\ \bibinfo {pages} {155106} (\bibinfo {year}
  {2015}{\natexlab{a}})}\BibitemShut {NoStop}%
\bibitem [{\citenamefont {Huynh}\ \emph {et~al.}(2014)\citenamefont {Huynh},
  \citenamefont {Tanabe}, \citenamefont {Urata}, \citenamefont {Oguro},
  \citenamefont {Heguri}, \citenamefont {Watanabe},\ and\ \citenamefont
  {Tanigaki}}]{Huynh14PRB}%
  \BibitemOpen
  \bibfield  {author} {\bibinfo {author} {\bibfnamefont {K.~K.}\ \bibnamefont
  {Huynh}}, \bibinfo {author} {\bibfnamefont {Y.}~\bibnamefont {Tanabe}},
  \bibinfo {author} {\bibfnamefont {T.}~\bibnamefont {Urata}}, \bibinfo
  {author} {\bibfnamefont {H.}~\bibnamefont {Oguro}}, \bibinfo {author}
  {\bibfnamefont {S.}~\bibnamefont {Heguri}}, \bibinfo {author} {\bibfnamefont
  {K.}~\bibnamefont {Watanabe}}, \ and\ \bibinfo {author} {\bibfnamefont
  {K.}~\bibnamefont {Tanigaki}},\ }\href {\doibase 10.1103/PhysRevB.90.144516}
  {\bibfield  {journal} {\bibinfo  {journal} {Phys. Rev. B}\ }\textbf {\bibinfo
  {volume} {90}},\ \bibinfo {pages} {144516} (\bibinfo {year}
  {2014})}\BibitemShut {NoStop}%
\bibitem [{\citenamefont {Watson}\ \emph
  {et~al.}(2015{\natexlab{b}})\citenamefont {Watson}, \citenamefont
  {Yamashita}, \citenamefont {Kasahara}, \citenamefont {Knafo}, \citenamefont
  {Nardone}, \citenamefont {B\'eard}, \citenamefont {Hardy}, \citenamefont
  {McCollam}, \citenamefont {Narayanan}, \citenamefont {Blake}, \citenamefont
  {Wolf}, \citenamefont {Haghighirad}, \citenamefont {Meingast}, \citenamefont
  {Schofield}, \citenamefont {v.~L\"ohneysen}, \citenamefont {Matsuda},
  \citenamefont {Coldea},\ and\ \citenamefont {Shibauchi}}]{Watson15PRL}%
  \BibitemOpen
  \bibfield  {author} {\bibinfo {author} {\bibfnamefont {M.~D.}\ \bibnamefont
  {Watson}}, \bibinfo {author} {\bibfnamefont {T.}~\bibnamefont {Yamashita}},
  \bibinfo {author} {\bibfnamefont {S.}~\bibnamefont {Kasahara}}, \bibinfo
  {author} {\bibfnamefont {W.}~\bibnamefont {Knafo}}, \bibinfo {author}
  {\bibfnamefont {M.}~\bibnamefont {Nardone}}, \bibinfo {author} {\bibfnamefont
  {J.}~\bibnamefont {B\'eard}}, \bibinfo {author} {\bibfnamefont
  {F.}~\bibnamefont {Hardy}}, \bibinfo {author} {\bibfnamefont
  {A.}~\bibnamefont {McCollam}}, \bibinfo {author} {\bibfnamefont
  {A.}~\bibnamefont {Narayanan}}, \bibinfo {author} {\bibfnamefont {S.~F.}\
  \bibnamefont {Blake}}, \bibinfo {author} {\bibfnamefont {T.}~\bibnamefont
  {Wolf}}, \bibinfo {author} {\bibfnamefont {A.~A.}\ \bibnamefont
  {Haghighirad}}, \bibinfo {author} {\bibfnamefont {C.}~\bibnamefont
  {Meingast}}, \bibinfo {author} {\bibfnamefont {A.~J.}\ \bibnamefont
  {Schofield}}, \bibinfo {author} {\bibfnamefont {H.}~\bibnamefont
  {v.~L\"ohneysen}}, \bibinfo {author} {\bibfnamefont {Y.}~\bibnamefont
  {Matsuda}}, \bibinfo {author} {\bibfnamefont {A.~I.}\ \bibnamefont {Coldea}},
  \ and\ \bibinfo {author} {\bibfnamefont {T.}~\bibnamefont {Shibauchi}},\
  }\href {\doibase 10.1103/PhysRevLett.115.027006} {\bibfield  {journal}
  {\bibinfo  {journal} {Phys. Rev. Lett.}\ }\textbf {\bibinfo {volume} {115}},\
  \bibinfo {pages} {027006} (\bibinfo {year} {2015}{\natexlab{b}})}\BibitemShut
  {NoStop}%
\bibitem [{\citenamefont {Zhang}\ \emph {et~al.}(2015)\citenamefont {Zhang},
  \citenamefont {Yi}, \citenamefont {Liu}, \citenamefont {Li}, \citenamefont
  {Lee}, \citenamefont {Moore}, \citenamefont {Hashimoto}, \citenamefont
  {Masamichi}, \citenamefont {H.~Eisaki~and}, \citenamefont {Hussain},
  \citenamefont {Devereaux}, \citenamefont {Shen},\ and\ \citenamefont
  {Lu}}]{Zhang15condmat}%
  \BibitemOpen
  \bibfield  {author} {\bibinfo {author} {\bibfnamefont {Y.}~\bibnamefont
  {Zhang}}, \bibinfo {author} {\bibfnamefont {M.}~\bibnamefont {Yi}}, \bibinfo
  {author} {\bibfnamefont {Z.-K.}\ \bibnamefont {Liu}}, \bibinfo {author}
  {\bibfnamefont {W.}~\bibnamefont {Li}}, \bibinfo {author} {\bibfnamefont
  {J.~J.}\ \bibnamefont {Lee}}, \bibinfo {author} {\bibfnamefont {R.~G.}\
  \bibnamefont {Moore}}, \bibinfo {author} {\bibfnamefont {M.}~\bibnamefont
  {Hashimoto}}, \bibinfo {author} {\bibfnamefont {N.}~\bibnamefont
  {Masamichi}}, \bibinfo {author} {\bibfnamefont {S.~K.~M.}\ \bibnamefont
  {H.~Eisaki~and}}, \bibinfo {author} {\bibfnamefont {Z.}~\bibnamefont
  {Hussain}}, \bibinfo {author} {\bibfnamefont {T.~P.}\ \bibnamefont
  {Devereaux}}, \bibinfo {author} {\bibfnamefont {Z.-X.}\ \bibnamefont {Shen}},
  \ and\ \bibinfo {author} {\bibfnamefont {D.~H.}\ \bibnamefont {Lu}},\
  }\href@noop {} {\bibfield  {journal} {\bibinfo  {journal} {arXiv:1503.01556}\
  } (\bibinfo {year} {2015})}\BibitemShut {NoStop}%
\bibitem [{\citenamefont {Randeria}\ and\ \citenamefont
  {Taylor}(2014)}]{Randeria14ARCMP}%
  \BibitemOpen
  \bibfield  {author} {\bibinfo {author} {\bibfnamefont {M.}~\bibnamefont
  {Randeria}}\ and\ \bibinfo {author} {\bibfnamefont {E.}~\bibnamefont
  {Taylor}},\ }\href {\doibase 10.1146/annurev-conmatphys-031113-133829}
  {\bibfield  {journal} {\bibinfo  {journal} {Annu. Rev. Condens. Matter
  Phys.}\ }\textbf {\bibinfo {volume} {5}},\ \bibinfo {pages} {209} (\bibinfo
  {year} {2014})}\BibitemShut {NoStop}%
\bibitem [{\citenamefont {Kasahara}\ \emph {et~al.}(2014)\citenamefont
  {Kasahara}, \citenamefont {Watashige}, \citenamefont {Hanaguri},
  \citenamefont {Kohsaka}, \citenamefont {Yamashita}, \citenamefont
  {Shimoyama}, \citenamefont {Mizukami}, \citenamefont {Endo}, \citenamefont
  {Ikeda}, \citenamefont {Aoyama}, \citenamefont {Terashima}, \citenamefont
  {Uji}, \citenamefont {Wolf}, \citenamefont {von L\"ohneysen}, \citenamefont
  {Shibauchi},\ and\ \citenamefont {Matsuda}}]{Kasahara14PNAS}%
  \BibitemOpen
  \bibfield  {author} {\bibinfo {author} {\bibfnamefont {S.}~\bibnamefont
  {Kasahara}}, \bibinfo {author} {\bibfnamefont {T.}~\bibnamefont {Watashige}},
  \bibinfo {author} {\bibfnamefont {T.}~\bibnamefont {Hanaguri}}, \bibinfo
  {author} {\bibfnamefont {Y.}~\bibnamefont {Kohsaka}}, \bibinfo {author}
  {\bibfnamefont {T.}~\bibnamefont {Yamashita}}, \bibinfo {author}
  {\bibfnamefont {Y.}~\bibnamefont {Shimoyama}}, \bibinfo {author}
  {\bibfnamefont {Y.}~\bibnamefont {Mizukami}}, \bibinfo {author}
  {\bibfnamefont {R.}~\bibnamefont {Endo}}, \bibinfo {author} {\bibfnamefont
  {H.}~\bibnamefont {Ikeda}}, \bibinfo {author} {\bibfnamefont
  {K.}~\bibnamefont {Aoyama}}, \bibinfo {author} {\bibfnamefont
  {T.}~\bibnamefont {Terashima}}, \bibinfo {author} {\bibfnamefont
  {S.}~\bibnamefont {Uji}}, \bibinfo {author} {\bibfnamefont {T.}~\bibnamefont
  {Wolf}}, \bibinfo {author} {\bibfnamefont {H.}~\bibnamefont {von
  L\"ohneysen}}, \bibinfo {author} {\bibfnamefont {T.}~\bibnamefont
  {Shibauchi}}, \ and\ \bibinfo {author} {\bibfnamefont {Y.}~\bibnamefont
  {Matsuda}},\ }\href@noop {} {\bibfield  {journal} {\bibinfo  {journal} {Proc.
  Natl. Acad. Sci. U. S. A.}\ }\textbf {\bibinfo {volume} {111}},\ \bibinfo
  {pages} {16309} (\bibinfo {year} {2014})}\BibitemShut {NoStop}%
\bibitem [{\citenamefont {B\"ohmer}\ \emph {et~al.}(2013)\citenamefont
  {B\"ohmer}, \citenamefont {Hardy}, \citenamefont {Eilers}, \citenamefont
  {Ernst}, \citenamefont {Adelmann}, \citenamefont {Schweiss}, \citenamefont
  {Wolf},\ and\ \citenamefont {Meingast}}]{Bohmer13PRB}%
  \BibitemOpen
  \bibfield  {author} {\bibinfo {author} {\bibfnamefont {A.~E.}\ \bibnamefont
  {B\"ohmer}}, \bibinfo {author} {\bibfnamefont {F.}~\bibnamefont {Hardy}},
  \bibinfo {author} {\bibfnamefont {F.}~\bibnamefont {Eilers}}, \bibinfo
  {author} {\bibfnamefont {D.}~\bibnamefont {Ernst}}, \bibinfo {author}
  {\bibfnamefont {P.}~\bibnamefont {Adelmann}}, \bibinfo {author}
  {\bibfnamefont {P.}~\bibnamefont {Schweiss}}, \bibinfo {author}
  {\bibfnamefont {T.}~\bibnamefont {Wolf}}, \ and\ \bibinfo {author}
  {\bibfnamefont {C.}~\bibnamefont {Meingast}},\ }\href {\doibase
  10.1103/PhysRevB.87.180505} {\bibfield  {journal} {\bibinfo  {journal} {Phys.
  Rev. B}\ }\textbf {\bibinfo {volume} {87}},\ \bibinfo {pages} {180505}
  (\bibinfo {year} {2013})}\BibitemShut {NoStop}%
\bibitem [{\citenamefont {Uwatoko}\ \emph {et~al.}(2002)\citenamefont
  {Uwatoko}, \citenamefont {Todo}, \citenamefont {Ueda}, \citenamefont
  {Uchida}, \citenamefont {Kosaka}, \citenamefont {Mori},\ and\ \citenamefont
  {Matsumoto}}]{Uwatoko02JPCM}%
  \BibitemOpen
  \bibfield  {author} {\bibinfo {author} {\bibfnamefont {Y.}~\bibnamefont
  {Uwatoko}}, \bibinfo {author} {\bibfnamefont {S.}~\bibnamefont {Todo}},
  \bibinfo {author} {\bibfnamefont {K.}~\bibnamefont {Ueda}}, \bibinfo {author}
  {\bibfnamefont {A.}~\bibnamefont {Uchida}}, \bibinfo {author} {\bibfnamefont
  {M.}~\bibnamefont {Kosaka}}, \bibinfo {author} {\bibfnamefont
  {N.}~\bibnamefont {Mori}}, \ and\ \bibinfo {author} {\bibfnamefont
  {T.}~\bibnamefont {Matsumoto}},\ }\href
  {http://stacks.iop.org/0953-8984/14/i=44/a=469} {\bibfield  {journal}
  {\bibinfo  {journal} {J. Phys.: Condens. Matter}\ }\textbf {\bibinfo {volume}
  {14}},\ \bibinfo {pages} {11291} (\bibinfo {year} {2002})}\BibitemShut
  {NoStop}%
\bibitem [{\citenamefont {Murata}\ \emph {et~al.}(2008)\citenamefont {Murata},
  \citenamefont {Yokogawa}, \citenamefont {Yoshino}, \citenamefont {Klotz},
  \citenamefont {Munsch}, \citenamefont {Irizawa}, \citenamefont {Nishiyama},
  \citenamefont {Iizuka}, \citenamefont {Nanba}, \citenamefont {Okada},
  \citenamefont {Shiraga},\ and\ \citenamefont {Aoyama}}]{Murata08RSI}%
  \BibitemOpen
  \bibfield  {author} {\bibinfo {author} {\bibfnamefont {K.}~\bibnamefont
  {Murata}}, \bibinfo {author} {\bibfnamefont {K.}~\bibnamefont {Yokogawa}},
  \bibinfo {author} {\bibfnamefont {H.}~\bibnamefont {Yoshino}}, \bibinfo
  {author} {\bibfnamefont {S.}~\bibnamefont {Klotz}}, \bibinfo {author}
  {\bibfnamefont {P.}~\bibnamefont {Munsch}}, \bibinfo {author} {\bibfnamefont
  {A.}~\bibnamefont {Irizawa}}, \bibinfo {author} {\bibfnamefont
  {M.}~\bibnamefont {Nishiyama}}, \bibinfo {author} {\bibfnamefont
  {K.}~\bibnamefont {Iizuka}}, \bibinfo {author} {\bibfnamefont
  {T.}~\bibnamefont {Nanba}}, \bibinfo {author} {\bibfnamefont
  {T.}~\bibnamefont {Okada}}, \bibinfo {author} {\bibfnamefont
  {Y.}~\bibnamefont {Shiraga}}, \ and\ \bibinfo {author} {\bibfnamefont
  {S.}~\bibnamefont {Aoyama}},\ }\href {\doibase 10.1063/1.2964117} {\bibfield
  {journal} {\bibinfo  {journal} {Rev. Sci. Instrum.}\ }\textbf {\bibinfo
  {volume} {79}},\ \bibinfo {eid} {085101} (\bibinfo {year}
  {2008})}\BibitemShut {NoStop}%
\bibitem [{\citenamefont {Kaluarachchi}\ \emph {et~al.}(2016)\citenamefont
  {Kaluarachchi}, \citenamefont {Taufour}, \citenamefont {B\"ohmer},
  \citenamefont {Tanatar}, \citenamefont {Bud'ko}, \citenamefont {Kogan},
  \citenamefont {Prozorov},\ and\ \citenamefont
  {Canfield}}]{Kaluarachchi16PRB}%
  \BibitemOpen
  \bibfield  {author} {\bibinfo {author} {\bibfnamefont {U.~S.}\ \bibnamefont
  {Kaluarachchi}}, \bibinfo {author} {\bibfnamefont {V.}~\bibnamefont
  {Taufour}}, \bibinfo {author} {\bibfnamefont {A.~E.}\ \bibnamefont
  {B\"ohmer}}, \bibinfo {author} {\bibfnamefont {M.~A.}\ \bibnamefont
  {Tanatar}}, \bibinfo {author} {\bibfnamefont {S.~L.}\ \bibnamefont {Bud'ko}},
  \bibinfo {author} {\bibfnamefont {V.~G.}\ \bibnamefont {Kogan}}, \bibinfo
  {author} {\bibfnamefont {R.}~\bibnamefont {Prozorov}}, \ and\ \bibinfo
  {author} {\bibfnamefont {P.~C.}\ \bibnamefont {Canfield}},\ }\href {\doibase
  10.1103/PhysRevB.93.064503} {\bibfield  {journal} {\bibinfo  {journal} {Phys.
  Rev. B}\ }\textbf {\bibinfo {volume} {93}},\ \bibinfo {pages} {064503}
  (\bibinfo {year} {2016})}\BibitemShut {NoStop}%
\bibitem [{\citenamefont {Press}\ \emph {et~al.}(2002)\citenamefont {Press},
  \citenamefont {Teukolsky}, \citenamefont {Vetterling},\ and\ \citenamefont
  {Flannery}}]{Press02}%
  \BibitemOpen
  \bibfield  {author} {\bibinfo {author} {\bibfnamefont {W.~H.}\ \bibnamefont
  {Press}}, \bibinfo {author} {\bibfnamefont {S.~A.}\ \bibnamefont
  {Teukolsky}}, \bibinfo {author} {\bibfnamefont {W.~T.}\ \bibnamefont
  {Vetterling}}, \ and\ \bibinfo {author} {\bibfnamefont {B.~P.}\ \bibnamefont
  {Flannery}},\ }\href@noop {} {\emph {\bibinfo {title} {Numerical Recipes in
  C++}}}\ (\bibinfo  {publisher} {Cambridge University Press},\ \bibinfo
  {address} {Cambridge},\ \bibinfo {year} {2002})\BibitemShut {NoStop}%
\bibitem [{\citenamefont {Ulrych}\ and\ \citenamefont
  {Bishop}(1975)}]{Ulrych75RGSP}%
  \BibitemOpen
  \bibfield  {author} {\bibinfo {author} {\bibfnamefont {T.~J.}\ \bibnamefont
  {Ulrych}}\ and\ \bibinfo {author} {\bibfnamefont {T.~N.}\ \bibnamefont
  {Bishop}},\ }\href@noop {} {\bibfield  {journal} {\bibinfo  {journal} {Rev.
  Geophys. Space Phys.}\ }\textbf {\bibinfo {volume} {13}},\ \bibinfo {pages}
  {183} (\bibinfo {year} {1975})}\BibitemShut {NoStop}%
\bibitem [{\citenamefont {Sigfusson}\ \emph {et~al.}(1992)\citenamefont
  {Sigfusson}, \citenamefont {Emilsson},\ and\ \citenamefont
  {Mattocks}}]{Sigfusson92PRB}%
  \BibitemOpen
  \bibfield  {author} {\bibinfo {author} {\bibfnamefont {T.~I.}\ \bibnamefont
  {Sigfusson}}, \bibinfo {author} {\bibfnamefont {K.~P.}\ \bibnamefont
  {Emilsson}}, \ and\ \bibinfo {author} {\bibfnamefont {P.}~\bibnamefont
  {Mattocks}},\ }\href {\doibase 10.1103/PhysRevB.46.10446} {\bibfield
  {journal} {\bibinfo  {journal} {Phys. Rev. B}\ }\textbf {\bibinfo {volume}
  {46}},\ \bibinfo {pages} {10446} (\bibinfo {year} {1992})}\BibitemShut
  {NoStop}%
\bibitem [{\citenamefont {Shoenberg}(1984)}]{Shoenberg84}%
  \BibitemOpen
  \bibfield  {author} {\bibinfo {author} {\bibfnamefont {D.}~\bibnamefont
  {Shoenberg}},\ }\href@noop {} {\emph {\bibinfo {title} {Magnetic oscillations
  in metals}}}\ (\bibinfo  {publisher} {Cambridge University Press},\ \bibinfo
  {address} {Cambridge},\ \bibinfo {year} {1984})\BibitemShut {NoStop}%
\bibitem [{\citenamefont {Shishido}\ \emph {et~al.}(2010)\citenamefont
  {Shishido}, \citenamefont {Bangura}, \citenamefont {Coldea}, \citenamefont
  {Tonegawa}, \citenamefont {Hashimoto}, \citenamefont {Kasahara},
  \citenamefont {Rourke}, \citenamefont {Ikeda}, \citenamefont {Terashima},
  \citenamefont {Settai}, \citenamefont {\ifmmode~\bar{O}\else \={O}\fi{}nuki},
  \citenamefont {Vignolles}, \citenamefont {Proust}, \citenamefont {Vignolle},
  \citenamefont {McCollam}, \citenamefont {Matsuda}, \citenamefont
  {Shibauchi},\ and\ \citenamefont {Carrington}}]{Shishido10PRL}%
  \BibitemOpen
  \bibfield  {author} {\bibinfo {author} {\bibfnamefont {H.}~\bibnamefont
  {Shishido}}, \bibinfo {author} {\bibfnamefont {A.~F.}\ \bibnamefont
  {Bangura}}, \bibinfo {author} {\bibfnamefont {A.~I.}\ \bibnamefont {Coldea}},
  \bibinfo {author} {\bibfnamefont {S.}~\bibnamefont {Tonegawa}}, \bibinfo
  {author} {\bibfnamefont {K.}~\bibnamefont {Hashimoto}}, \bibinfo {author}
  {\bibfnamefont {S.}~\bibnamefont {Kasahara}}, \bibinfo {author}
  {\bibfnamefont {P.~M.~C.}\ \bibnamefont {Rourke}}, \bibinfo {author}
  {\bibfnamefont {H.}~\bibnamefont {Ikeda}}, \bibinfo {author} {\bibfnamefont
  {T.}~\bibnamefont {Terashima}}, \bibinfo {author} {\bibfnamefont
  {R.}~\bibnamefont {Settai}}, \bibinfo {author} {\bibfnamefont
  {Y.}~\bibnamefont {\ifmmode~\bar{O}\else \={O}\fi{}nuki}}, \bibinfo {author}
  {\bibfnamefont {D.}~\bibnamefont {Vignolles}}, \bibinfo {author}
  {\bibfnamefont {C.}~\bibnamefont {Proust}}, \bibinfo {author} {\bibfnamefont
  {B.}~\bibnamefont {Vignolle}}, \bibinfo {author} {\bibfnamefont
  {A.}~\bibnamefont {McCollam}}, \bibinfo {author} {\bibfnamefont
  {Y.}~\bibnamefont {Matsuda}}, \bibinfo {author} {\bibfnamefont
  {T.}~\bibnamefont {Shibauchi}}, \ and\ \bibinfo {author} {\bibfnamefont
  {A.}~\bibnamefont {Carrington}},\ }\href {\doibase
  10.1103/PhysRevLett.104.057008} {\bibfield  {journal} {\bibinfo  {journal}
  {Phys. Rev. Lett.}\ }\textbf {\bibinfo {volume} {104}},\ \bibinfo {pages}
  {057008} (\bibinfo {year} {2010})}\BibitemShut {NoStop}%
\bibitem [{\citenamefont {Hashimoto}\ \emph {et~al.}(2012)\citenamefont
  {Hashimoto}, \citenamefont {Cho}, \citenamefont {Shibauchi}, \citenamefont
  {Kasahara}, \citenamefont {Mizukami}, \citenamefont {Katsumata},
  \citenamefont {Tsuruhara}, \citenamefont {Terashima}, \citenamefont {Ikeda},
  \citenamefont {Tanatar}, \citenamefont {Kitano}, \citenamefont {Salovich},
  \citenamefont {Giannetta}, \citenamefont {Walmsley}, \citenamefont
  {Carrington}, \citenamefont {Prozorov},\ and\ \citenamefont
  {Matsuda}}]{Hashimoto12Sci}%
  \BibitemOpen
  \bibfield  {author} {\bibinfo {author} {\bibfnamefont {K.}~\bibnamefont
  {Hashimoto}}, \bibinfo {author} {\bibfnamefont {K.}~\bibnamefont {Cho}},
  \bibinfo {author} {\bibfnamefont {T.}~\bibnamefont {Shibauchi}}, \bibinfo
  {author} {\bibfnamefont {S.}~\bibnamefont {Kasahara}}, \bibinfo {author}
  {\bibfnamefont {Y.}~\bibnamefont {Mizukami}}, \bibinfo {author}
  {\bibfnamefont {R.}~\bibnamefont {Katsumata}}, \bibinfo {author}
  {\bibfnamefont {Y.}~\bibnamefont {Tsuruhara}}, \bibinfo {author}
  {\bibfnamefont {T.}~\bibnamefont {Terashima}}, \bibinfo {author}
  {\bibfnamefont {H.}~\bibnamefont {Ikeda}}, \bibinfo {author} {\bibfnamefont
  {M.~A.}\ \bibnamefont {Tanatar}}, \bibinfo {author} {\bibfnamefont
  {H.}~\bibnamefont {Kitano}}, \bibinfo {author} {\bibfnamefont
  {N.}~\bibnamefont {Salovich}}, \bibinfo {author} {\bibfnamefont {R.~W.}\
  \bibnamefont {Giannetta}}, \bibinfo {author} {\bibfnamefont {P.}~\bibnamefont
  {Walmsley}}, \bibinfo {author} {\bibfnamefont {A.}~\bibnamefont
  {Carrington}}, \bibinfo {author} {\bibfnamefont {R.}~\bibnamefont
  {Prozorov}}, \ and\ \bibinfo {author} {\bibfnamefont {Y.}~\bibnamefont
  {Matsuda}},\ }\href {\doibase 10.1126/science.1219821} {\bibfield  {journal}
  {\bibinfo  {journal} {Science}\ }\textbf {\bibinfo {volume} {336}},\ \bibinfo
  {pages} {1554} (\bibinfo {year} {2012})}\BibitemShut {NoStop}%
\bibitem [{\citenamefont {Walmsley}\ \emph {et~al.}(2013)\citenamefont
  {Walmsley}, \citenamefont {Putzke}, \citenamefont {Malone}, \citenamefont
  {Guillam\'on}, \citenamefont {Vignolles}, \citenamefont {Proust},
  \citenamefont {Badoux}, \citenamefont {Coldea}, \citenamefont {Watson},
  \citenamefont {Kasahara}, \citenamefont {Mizukami}, \citenamefont
  {Shibauchi}, \citenamefont {Matsuda},\ and\ \citenamefont
  {Carrington}}]{Walmsley13PRL}%
  \BibitemOpen
  \bibfield  {author} {\bibinfo {author} {\bibfnamefont {P.}~\bibnamefont
  {Walmsley}}, \bibinfo {author} {\bibfnamefont {C.}~\bibnamefont {Putzke}},
  \bibinfo {author} {\bibfnamefont {L.}~\bibnamefont {Malone}}, \bibinfo
  {author} {\bibfnamefont {I.}~\bibnamefont {Guillam\'on}}, \bibinfo {author}
  {\bibfnamefont {D.}~\bibnamefont {Vignolles}}, \bibinfo {author}
  {\bibfnamefont {C.}~\bibnamefont {Proust}}, \bibinfo {author} {\bibfnamefont
  {S.}~\bibnamefont {Badoux}}, \bibinfo {author} {\bibfnamefont {A.~I.}\
  \bibnamefont {Coldea}}, \bibinfo {author} {\bibfnamefont {M.~D.}\
  \bibnamefont {Watson}}, \bibinfo {author} {\bibfnamefont {S.}~\bibnamefont
  {Kasahara}}, \bibinfo {author} {\bibfnamefont {Y.}~\bibnamefont {Mizukami}},
  \bibinfo {author} {\bibfnamefont {T.}~\bibnamefont {Shibauchi}}, \bibinfo
  {author} {\bibfnamefont {Y.}~\bibnamefont {Matsuda}}, \ and\ \bibinfo
  {author} {\bibfnamefont {A.}~\bibnamefont {Carrington}},\ }\href {\doibase
  10.1103/PhysRevLett.110.257002} {\bibfield  {journal} {\bibinfo  {journal}
  {Phys. Rev. Lett.}\ }\textbf {\bibinfo {volume} {110}},\ \bibinfo {pages}
  {257002} (\bibinfo {year} {2013})}\BibitemShut {NoStop}%
\bibitem [{\citenamefont {Shibauchi}\ \emph {et~al.}(2014)\citenamefont
  {Shibauchi}, \citenamefont {Carrington},\ and\ \citenamefont
  {Matsuda}}]{Shibauchi14ARCMP}%
  \BibitemOpen
  \bibfield  {author} {\bibinfo {author} {\bibfnamefont {T.}~\bibnamefont
  {Shibauchi}}, \bibinfo {author} {\bibfnamefont {A.}~\bibnamefont
  {Carrington}}, \ and\ \bibinfo {author} {\bibfnamefont {Y.}~\bibnamefont
  {Matsuda}},\ }\href {\doibase 10.1146/annurev-conmatphys-031113-133921}
  {\bibfield  {journal} {\bibinfo  {journal} {Annu. Rev. Condens. Matter
  Phys.}\ }\textbf {\bibinfo {volume} {5}},\ \bibinfo {pages} {113} (\bibinfo
  {year} {2014})}\BibitemShut {NoStop}%
\bibitem [{\citenamefont {Klotz}\ \emph {et~al.}(2009)\citenamefont {Klotz},
  \citenamefont {Chervin}, \citenamefont {Munsch},\ and\ \citenamefont
  {Marchand}}]{Klotz09JPhysD}%
  \BibitemOpen
  \bibfield  {author} {\bibinfo {author} {\bibfnamefont {S.}~\bibnamefont
  {Klotz}}, \bibinfo {author} {\bibfnamefont {J.-C.}\ \bibnamefont {Chervin}},
  \bibinfo {author} {\bibfnamefont {P.}~\bibnamefont {Munsch}}, \ and\ \bibinfo
  {author} {\bibfnamefont {G.~L.}\ \bibnamefont {Marchand}},\ }\href
  {http://stacks.iop.org/0022-3727/42/i=7/a=075413} {\bibfield  {journal}
  {\bibinfo  {journal} {J. Phys. D: Appl. Phys.}\ }\textbf {\bibinfo {volume}
  {42}},\ \bibinfo {pages} {075413} (\bibinfo {year} {2009})}\BibitemShut
  {NoStop}%
\bibitem [{\citenamefont {Tateiwa}\ and\ \citenamefont
  {Haga}(2009)}]{Tateiwa09RSI}%
  \BibitemOpen
  \bibfield  {author} {\bibinfo {author} {\bibfnamefont {N.}~\bibnamefont
  {Tateiwa}}\ and\ \bibinfo {author} {\bibfnamefont {Y.}~\bibnamefont {Haga}},\
  }\href
  {http://scitation.aip.org/content/aip/journal/rsi/80/12/10.1063/1.3265992}
  {\bibfield  {journal} {\bibinfo  {journal} {Rev. Sci. Instrum.}\ }\textbf
  {\bibinfo {volume} {80}},\ \bibinfo {eid} {123901} (\bibinfo {year}
  {2009})}\BibitemShut {NoStop}%
\bibitem [{\citenamefont {Terashima}\ \emph {et~al.}(2007)\citenamefont
  {Terashima}, \citenamefont {Takahide}, \citenamefont {Matsumoto},
  \citenamefont {Uji}, \citenamefont {Kimura}, \citenamefont {Aoki},\ and\
  \citenamefont {Harima}}]{Terashima07PRB}%
  \BibitemOpen
  \bibfield  {author} {\bibinfo {author} {\bibfnamefont {T.}~\bibnamefont
  {Terashima}}, \bibinfo {author} {\bibfnamefont {Y.}~\bibnamefont {Takahide}},
  \bibinfo {author} {\bibfnamefont {T.}~\bibnamefont {Matsumoto}}, \bibinfo
  {author} {\bibfnamefont {S.}~\bibnamefont {Uji}}, \bibinfo {author}
  {\bibfnamefont {N.}~\bibnamefont {Kimura}}, \bibinfo {author} {\bibfnamefont
  {H.}~\bibnamefont {Aoki}}, \ and\ \bibinfo {author} {\bibfnamefont
  {H.}~\bibnamefont {Harima}},\ }\href {\doibase 10.1103/PhysRevB.76.054506}
  {\bibfield  {journal} {\bibinfo  {journal} {Phys. Rev. B}\ }\textbf {\bibinfo
  {volume} {76}},\ \bibinfo {pages} {054506} (\bibinfo {year}
  {2007})}\BibitemShut {NoStop}%
\bibitem [{\citenamefont {Terashima}\ \emph
  {et~al.}(2014{\natexlab{b}})\citenamefont {Terashima}, \citenamefont {Kihou},
  \citenamefont {Sugii}, \citenamefont {Kikugawa}, \citenamefont {Matsumoto},
  \citenamefont {Ishida}, \citenamefont {Lee}, \citenamefont {Iyo},
  \citenamefont {Eisaki},\ and\ \citenamefont {Uji}}]{Terashima14PRB_KFA}%
  \BibitemOpen
  \bibfield  {author} {\bibinfo {author} {\bibfnamefont {T.}~\bibnamefont
  {Terashima}}, \bibinfo {author} {\bibfnamefont {K.}~\bibnamefont {Kihou}},
  \bibinfo {author} {\bibfnamefont {K.}~\bibnamefont {Sugii}}, \bibinfo
  {author} {\bibfnamefont {N.}~\bibnamefont {Kikugawa}}, \bibinfo {author}
  {\bibfnamefont {T.}~\bibnamefont {Matsumoto}}, \bibinfo {author}
  {\bibfnamefont {S.}~\bibnamefont {Ishida}}, \bibinfo {author} {\bibfnamefont
  {C.-H.}\ \bibnamefont {Lee}}, \bibinfo {author} {\bibfnamefont
  {A.}~\bibnamefont {Iyo}}, \bibinfo {author} {\bibfnamefont {H.}~\bibnamefont
  {Eisaki}}, \ and\ \bibinfo {author} {\bibfnamefont {S.}~\bibnamefont {Uji}},\
  }\href {\doibase 10.1103/PhysRevB.89.134520} {\bibfield  {journal} {\bibinfo
  {journal} {Phys. Rev. B}\ }\textbf {\bibinfo {volume} {89}},\ \bibinfo
  {pages} {134520} (\bibinfo {year} {2014}{\natexlab{b}})}\BibitemShut
  {NoStop}%
\bibitem [{\citenamefont {Dai}\ \emph {et~al.}(2012)\citenamefont {Dai},
  \citenamefont {Hu},\ and\ \citenamefont {Dagotto}}]{Dai12NatPhys}%
  \BibitemOpen
  \bibfield  {author} {\bibinfo {author} {\bibfnamefont {P.}~\bibnamefont
  {Dai}}, \bibinfo {author} {\bibfnamefont {J.}~\bibnamefont {Hu}}, \ and\
  \bibinfo {author} {\bibfnamefont {E.}~\bibnamefont {Dagotto}},\ }\href
  {http://dx.doi.org/10.1038/nphys2438} {\bibfield  {journal} {\bibinfo
  {journal} {Nat. Phys.}\ }\textbf {\bibinfo {volume} {8}},\ \bibinfo {pages}
  {709} (\bibinfo {year} {2012})}\BibitemShut {NoStop}%
\bibitem [{\citenamefont {Johannes}\ and\ \citenamefont
  {Mazin}(2009)}]{Johannes09PRB}%
  \BibitemOpen
  \bibfield  {author} {\bibinfo {author} {\bibfnamefont {M.~D.}\ \bibnamefont
  {Johannes}}\ and\ \bibinfo {author} {\bibfnamefont {I.~I.}\ \bibnamefont
  {Mazin}},\ }\href {\doibase 10.1103/PhysRevB.79.220510} {\bibfield  {journal}
  {\bibinfo  {journal} {Phys. Rev. B}\ }\textbf {\bibinfo {volume} {79}},\
  \bibinfo {pages} {220510} (\bibinfo {year} {2009})}\BibitemShut {NoStop}%
\bibitem [{\citenamefont {Heil}\ \emph {et~al.}(2014)\citenamefont {Heil},
  \citenamefont {Sormann}, \citenamefont {Boeri}, \citenamefont {Aichhorn},\
  and\ \citenamefont {von~der Linden}}]{Heil14PRB}%
  \BibitemOpen
  \bibfield  {author} {\bibinfo {author} {\bibfnamefont {C.}~\bibnamefont
  {Heil}}, \bibinfo {author} {\bibfnamefont {H.}~\bibnamefont {Sormann}},
  \bibinfo {author} {\bibfnamefont {L.}~\bibnamefont {Boeri}}, \bibinfo
  {author} {\bibfnamefont {M.}~\bibnamefont {Aichhorn}}, \ and\ \bibinfo
  {author} {\bibfnamefont {W.}~\bibnamefont {von~der Linden}},\ }\href
  {\doibase 10.1103/PhysRevB.90.115143} {\bibfield  {journal} {\bibinfo
  {journal} {Phys. Rev. B}\ }\textbf {\bibinfo {volume} {90}},\ \bibinfo
  {pages} {115143} (\bibinfo {year} {2014})}\BibitemShut {NoStop}%
\bibitem [{\citenamefont {Rotter}\ \emph
  {et~al.}(2008{\natexlab{b}})\citenamefont {Rotter}, \citenamefont {Tegel},
  \citenamefont {Johrendt}, \citenamefont {Schellenberg}, \citenamefont
  {Hermes},\ and\ \citenamefont {P\"{o}ttgen}}]{Rotter08PRB}%
  \BibitemOpen
  \bibfield  {author} {\bibinfo {author} {\bibfnamefont {M.}~\bibnamefont
  {Rotter}}, \bibinfo {author} {\bibfnamefont {M.}~\bibnamefont {Tegel}},
  \bibinfo {author} {\bibfnamefont {D.}~\bibnamefont {Johrendt}}, \bibinfo
  {author} {\bibfnamefont {I.}~\bibnamefont {Schellenberg}}, \bibinfo {author}
  {\bibfnamefont {W.}~\bibnamefont {Hermes}}, \ and\ \bibinfo {author}
  {\bibfnamefont {R.}~\bibnamefont {P\"{o}ttgen}},\ }\href {\doibase
  10.1103/PhysRevB.78.020503} {\bibfield  {journal} {\bibinfo  {journal} {Phys.
  Rev. B}\ }\textbf {\bibinfo {volume} {78}},\ \bibinfo {eid} {020503}
  (\bibinfo {year} {2008}{\natexlab{b}})}\BibitemShut {NoStop}%
\bibitem [{\citenamefont {Nimori}\ \emph {et~al.}(1993)\citenamefont {Nimori},
  \citenamefont {Kido}, \citenamefont {Kido}, \citenamefont {Nakagawa},
  \citenamefont {Haga},\ and\ \citenamefont {Suzuki}}]{Nimori93PhysB}%
  \BibitemOpen
  \bibfield  {author} {\bibinfo {author} {\bibfnamefont {S.}~\bibnamefont
  {Nimori}}, \bibinfo {author} {\bibfnamefont {A.}~\bibnamefont {Kido}},
  \bibinfo {author} {\bibfnamefont {G.}~\bibnamefont {Kido}}, \bibinfo {author}
  {\bibfnamefont {Y.}~\bibnamefont {Nakagawa}}, \bibinfo {author}
  {\bibfnamefont {Y.}~\bibnamefont {Haga}}, \ and\ \bibinfo {author}
  {\bibfnamefont {T.}~\bibnamefont {Suzuki}},\ }\href@noop {} {\bibfield
  {journal} {\bibinfo  {journal} {Physica B}\ }\textbf {\bibinfo {volume}
  {186--188}},\ \bibinfo {pages} {173 } (\bibinfo {year} {1993})}\BibitemShut
  {NoStop}%
\bibitem [{\citenamefont {Harrison}\ \emph {et~al.}(1993)\citenamefont
  {Harrison}, \citenamefont {Meeson}, \citenamefont {Probst},\ and\
  \citenamefont {Springford}}]{Harrison93JPCM}%
  \BibitemOpen
  \bibfield  {author} {\bibinfo {author} {\bibfnamefont {N.}~\bibnamefont
  {Harrison}}, \bibinfo {author} {\bibfnamefont {P.}~\bibnamefont {Meeson}},
  \bibinfo {author} {\bibfnamefont {P.~A.}\ \bibnamefont {Probst}}, \ and\
  \bibinfo {author} {\bibfnamefont {M.}~\bibnamefont {Springford}},\ }\href
  {http://stacks.iop.org/0953-8984/5/i=40/a=018} {\bibfield  {journal}
  {\bibinfo  {journal} {J. Phys.: Condens. Matter}\ }\textbf {\bibinfo {volume}
  {5}},\ \bibinfo {pages} {7435} (\bibinfo {year} {1993})}\BibitemShut
  {NoStop}%
\bibitem [{\citenamefont {Uji}\ \emph {et~al.}(1995)\citenamefont {Uji},
  \citenamefont {Terashima}, \citenamefont {Aoki}, \citenamefont {Brooks},
  \citenamefont {Tokumoto}, \citenamefont {Kinoshita}, \citenamefont
  {Kinoshita}, \citenamefont {Tanaka},\ and\ \citenamefont
  {Anzai}}]{Uji95SynthM}%
  \BibitemOpen
  \bibfield  {author} {\bibinfo {author} {\bibfnamefont {S.}~\bibnamefont
  {Uji}}, \bibinfo {author} {\bibfnamefont {T.}~\bibnamefont {Terashima}},
  \bibinfo {author} {\bibfnamefont {H.}~\bibnamefont {Aoki}}, \bibinfo {author}
  {\bibfnamefont {J.}~\bibnamefont {Brooks}}, \bibinfo {author} {\bibfnamefont
  {M.}~\bibnamefont {Tokumoto}}, \bibinfo {author} {\bibfnamefont
  {N.}~\bibnamefont {Kinoshita}}, \bibinfo {author} {\bibfnamefont
  {T.}~\bibnamefont {Kinoshita}}, \bibinfo {author} {\bibfnamefont
  {Y.}~\bibnamefont {Tanaka}}, \ and\ \bibinfo {author} {\bibfnamefont
  {H.}~\bibnamefont {Anzai}},\ }\href {\doibase
  http://dx.doi.org/10.1016/0379-6779(94)02659-M} {\bibfield  {journal}
  {\bibinfo  {journal} {Synth. Met.}\ }\textbf {\bibinfo {volume} {70}},\
  \bibinfo {pages} {807 } (\bibinfo {year} {1995})}\BibitemShut {NoStop}%
\bibitem [{\citenamefont {Terashima}\ \emph {et~al.}(1998)\citenamefont
  {Terashima}, \citenamefont {Haworth}, \citenamefont {Takashita},
  \citenamefont {Uji}, \citenamefont {Aoki}, \citenamefont {Haga},
  \citenamefont {Uesawa},\ and\ \citenamefont {Suzuki}}]{Terashima98JMMM}%
  \BibitemOpen
  \bibfield  {author} {\bibinfo {author} {\bibfnamefont {T.}~\bibnamefont
  {Terashima}}, \bibinfo {author} {\bibfnamefont {C.}~\bibnamefont {Haworth}},
  \bibinfo {author} {\bibfnamefont {M.}~\bibnamefont {Takashita}}, \bibinfo
  {author} {\bibfnamefont {S.}~\bibnamefont {Uji}}, \bibinfo {author}
  {\bibfnamefont {H.}~\bibnamefont {Aoki}}, \bibinfo {author} {\bibfnamefont
  {Y.}~\bibnamefont {Haga}}, \bibinfo {author} {\bibfnamefont {A.}~\bibnamefont
  {Uesawa}}, \ and\ \bibinfo {author} {\bibfnamefont {T.}~\bibnamefont
  {Suzuki}},\ }\href {\doibase http://dx.doi.org/10.1016/S0304-8853(97)00368-5}
  {\bibfield  {journal} {\bibinfo  {journal} {J. Magn. Magn. Mater.}\ }\textbf
  {\bibinfo {volume} {177--181}},\ \bibinfo {pages} {421 } (\bibinfo {year}
  {1998})}\BibitemShut {NoStop}%
\bibitem [{\citenamefont {Coldea}\ \emph {et~al.}(2004)\citenamefont {Coldea},
  \citenamefont {Bangura}, \citenamefont {Singleton}, \citenamefont {Ardavan},
  \citenamefont {Akutsu-Sato}, \citenamefont {Akutsu}, \citenamefont {Turner},\
  and\ \citenamefont {Day}}]{Coldea04PRB}%
  \BibitemOpen
  \bibfield  {author} {\bibinfo {author} {\bibfnamefont {A.~I.}\ \bibnamefont
  {Coldea}}, \bibinfo {author} {\bibfnamefont {A.~F.}\ \bibnamefont {Bangura}},
  \bibinfo {author} {\bibfnamefont {J.}~\bibnamefont {Singleton}}, \bibinfo
  {author} {\bibfnamefont {A.}~\bibnamefont {Ardavan}}, \bibinfo {author}
  {\bibfnamefont {A.}~\bibnamefont {Akutsu-Sato}}, \bibinfo {author}
  {\bibfnamefont {H.}~\bibnamefont {Akutsu}}, \bibinfo {author} {\bibfnamefont
  {S.~S.}\ \bibnamefont {Turner}}, \ and\ \bibinfo {author} {\bibfnamefont
  {P.}~\bibnamefont {Day}},\ }\href {\doibase 10.1103/PhysRevB.69.085112}
  {\bibfield  {journal} {\bibinfo  {journal} {Phys. Rev. B}\ }\textbf {\bibinfo
  {volume} {69}},\ \bibinfo {pages} {085112} (\bibinfo {year}
  {2004})}\BibitemShut {NoStop}%
\bibitem [{\citenamefont {Kang}\ and\ \citenamefont {Chung}(2009)}]{Kang09PRB}%
  \BibitemOpen
  \bibfield  {author} {\bibinfo {author} {\bibfnamefont {W.}~\bibnamefont
  {Kang}}\ and\ \bibinfo {author} {\bibfnamefont {O.-H.}\ \bibnamefont
  {Chung}},\ }\href {\doibase 10.1103/PhysRevB.79.045115} {\bibfield  {journal}
  {\bibinfo  {journal} {Phys. Rev. B}\ }\textbf {\bibinfo {volume} {79}},\
  \bibinfo {pages} {045115} (\bibinfo {year} {2009})}\BibitemShut {NoStop}%
\bibitem [{\citenamefont {Pardo-Ig\'uzquiza}\ and\ \citenamefont
  {Rodr\'{\i}guez-Tovar}(2005)}]{Pardo05ComputGeosci}%
  \BibitemOpen
  \bibfield  {author} {\bibinfo {author} {\bibfnamefont {E.}~\bibnamefont
  {Pardo-Ig\'uzquiza}}\ and\ \bibinfo {author} {\bibfnamefont {F.}~\bibnamefont
  {Rodr\'{\i}guez-Tovar}},\ }\href@noop {} {\bibfield  {journal} {\bibinfo
  {journal} {Comput. Geosci.}\ }\textbf {\bibinfo {volume} {31}},\ \bibinfo
  {pages} {555} (\bibinfo {year} {2005})}\BibitemShut {NoStop}%
\end{thebibliography}
\end{document}